\documentclass[fleqn,usenatbib]{mnras}

\usepackage{mathptmx}
\usepackage[T1]{fontenc}
\usepackage{ae,aecompl}

\usepackage{graphicx}	% Including figure files
\usepackage{amsmath}	% Advanced maths commands
\usepackage{amssymb}	% Extra maths symbols
\usepackage{hyperref}
\usepackage{subfigure}
\usepackage{color}
\usepackage{gensymb}
\usepackage[toc,page]{appendix}

\title[Continuous gravitational wave]{Continuous gravitational wave from magnetized white dwarfs and neutron stars: possible missions for LISA, DECIGO, BBO, ET detectors}

\author[S. Kalita \& B. Mukhopadhyay]{
Surajit Kalita\thanks{E-mail: surajitk@iisc.ac.in} and
Banibrata Mukhopadhyay\thanks{E-mail: bm@iisc.ac.in}
\\
Department of Physics, Indian Institute of Science, Bangalore 560012, India
}

\date{Accepted XXX. Received YYY; in original form ZZZ}

\pubyear{2019}

\begin{document}
\label{firstpage}
\pagerange{\pageref{firstpage}--\pageref{lastpage}}
\maketitle

% Abstract of the paper
\begin{abstract}
Recent detection of gravitational wave from nine black hole merger events and one neutron star merger event by LIGO and VIRGO shed a new light in the field of astrophysics. On the other hand, in the past decade, a few super-Chandrasekhar white dwarf candidates have been inferred through the peak luminosity of the light-curves of a few peculiar type Ia supernovae, though there is no direct detection of these objects so far. Similarly, a number of neutron stars with mass $>2M_\odot$ have also been observed. Continuous gravitational wave can be one of the alternate ways to detect these compact objects directly. It was already argued that magnetic field is one of the prominent physics to form super-Chandrasekhar white dwarfs and massive neutron stars. If such compact objects are rotating with certain angular frequency, then they can efficiently emit gravitational radiation, provided their magnetic field and rotation axes are not aligned, and these gravitational waves can be detected by some of the upcoming detectors, e.g. LISA, BBO, DECIGO, Einstein Telescope etc. This will certainly be a direct detection of rotating magnetized white dwarfs as well as massive neutron stars.
\end{abstract}

% Select between one and six entries from the list of approved keywords.
% Don't make up new ones.

%%%%%%%%%%%%%%%%%%%%%%%%%%%%%%%%%%%%%%%%%%%%%%%%%%

%%%%%%%%%%%%%%%%% BODY OF PAPER %%%%%%%%%%%%%%%%%%

\section{Introduction}\label{Introduction}
Over the past 100 years, Einstein's theory of general relativity is the most efficient theory to understand theory of gravity. It can easily explain the physics of strong gravity around various compact sources such as black holes, neutron stars, white dwarfs etc. Moreover, general relativity is the backbone to understand the various eras of cosmology after the Big Bang. This theory has already been well tested through various experiments ranging from deflection of light rays in strong gravity, perihelion precession of Mercury, gravitational red-shift of light etc. More recently, its another important consequence has been confirmed through the detection of the gravitational wave by LIGO. Gravitational wave is the ripple in spacetime, formed due to distortion in the curvature of the spacetime and propagates at the speed of light \citep{1973grav.book.....M}. The event of two black hole mergers, named as GW150914, was the first to confirm directly the existence of gravitational wave \citep{2016PhRvL.116f1102A}. This is the event in which two black holes of masses $\sim 35.4 M_\odot$ and $\sim 29.8 M_\odot$ merge together to form a bigger black hole of mass $\sim 62.2 M_\odot$. Thereafter, 9 more such events have been observed which confirm the existence of gravitational wave \citep{2016PhRvL.116x1103A, 2017PhRvL.118v1101A, 2017ApJ...851L..35A, 2017PhRvL.119n1101A, 2017PhRvL.119p1101A,2018arXiv181112940T}. Among them, the event GW170817 confirms the merger of two neutron stars to form a stellar mass black hole \citep{2017PhRvL.119p1101A}.

Gravitational wave is emitted if the system has a non-zero quadrupole moment \citep{1973grav.book.....M}. The binary systems of all the above-mentioned events, which were detected by LIGO/VIRGO, possess non-zero quadrupole moment during the period of the merger. However, single spinning massive objects may also be able to emit gravitational wave, provided the object should have a non-zero quadrupole moment. This type of gravitational wave is known as the continuous gravitational wave because it is continuously emitted at certain frequency and amplitude \citep{1979PhRvD..20..351Z}. Different possibilities of generation of continuous gravitational wave have already been proposed in various literature, such as, sources with breaking of axisymmetry through misalignment of magnetic field and rotation axes \citep{1996A&A...312..675B, 2002MNRAS.331..203J,2010CQGra..27s4006F,2017MNRAS.467.4484F,2017MNRAS.472.3564M,2017ApJ...844..112G}, presence of mountains at the stellar surface \citep{2005Ap.....48...53S,2008MNRAS.385..531H,2009PhRvL.102s1102H,2011CQGra..28k4014G,2018ASSL..457..673G}, accreting neutron stars \citep{1998ApJ...501L..89B,2000MNRAS.319..902U,2008MNRAS.389..839W,2018PhRvL.121m1105T} etc. One may find more on continuous gravitational wave in the recent review by \cite{2017MPLA...3230035R}. In this paper, we show that continuous gravitational wave can be emitted from rotating magnetized white dwarfs, namely B-WDs, and will possibly be detected by the upcoming gravitational wave detectors such as LISA, DECIGO, BBO etc. We also argue that white dwarfs in a binary system emit much stronger gravitational wave which can also be detected by these detectors, but at a different frequency range. Moreover, we show that strong gravitational radiation can be emitted from rotating magnetized neutron stars, namely B-NSs, which can be in fact detected by Einstein telescope (ET). Eventually we argue how to constrain magnetic field of white dwarfs/neutron stars from gravitational wave detection.

A white dwarf is the end state of a star with mass $\lesssim 8 M_\odot$. In a white dwarf, the inward gravitational force is balanced by the force due to outward electron degeneracy pressure. If a white dwarf has a binary partner, it starts pulling matter out from the partner due to its high gravity resulting in the increase of mass of the white dwarf. When it gains sufficient amount of matter, beyond a certain mass, known as the Chandrasekhar mass-limit: currently accepted value $\sim 1.4M_\odot$ for a carbon-oxygen non-magnetized and non-rotating white dwarf \citep{1931ApJ....74...81C}, this pressure balance is no longer sustained and it burns out to produce type Ia supernova (SNeIa) with extremely high luminosity. Nevertheless, recent observations have suggested several peculiar over-luminous SNeIa \citep{2006Natur.443..308H, 2010ApJ...713.1073S}, which are believed to be originating from white dwarfs of super-Chandrasekhar mass as high as 2.8$M_\odot$. Ostriker and his collaborators first showed that rotation (and also magnetic field) of the white dwarf can lead to the violation of the Chandrasekhar mass-limit \citep{1968ApJ...153..797O}, but they could not reveal any limiting mass. More recently, magnetic field \citep{2013PhRvL.110g1102D, 2015JCAP...05..016D, 2015MNRAS.454..752S}, modified theories of gravity \citep{2015JCAP...05..045D, 2018JCAP...09..007K, 2017EPJC...77..871C}, generalized Heisenberg uncertainty principle \citep{2018JCAP...09..015O} etc. have been proposed as some of the prominent possibilities to explain super-Chandrasekhar white dwarfs and also corresponding limiting mass. Moreover, in case of neutron stars, which are the end state of stars with masses between $8M_\odot <M \lesssim 20M_\odot$, a few observations suggest that it may have mass $>2M_\odot$ \citep{2018ApJ...859...54L, 0004-637X-728-2-95}. These high mass neutron stars can also be inferred to be formed due to magnetic field and rotation along with steep equation of state \citep{2014MNRAS.439.3541P}.

In this paper, we show that if the rotation and the magnetic field axes are not aligned to each other, rotating B-WDs and B-NSs can be prominent sources for generation of continuous gravitational wave which can be detected by LISA, DECIGO, BBO, ET etc. In section \ref{methodology}, we illustrate the model of the compact object which we consider to solve the problem. Subsequently in section \ref{Result}, we discuss our results for B-WDs and B-NSs considering various central densities and magnetic field geometries with the change of angular frequency. We also show that a system of white dwarfs including B-WD and its binary companion, having a non-zero quadrupole moment, generates significant gravitational radiation. In section \ref{gwem}, we compare the luminosity due to gravitational radiation with the electromagnetic counterpart of these magnetized objects. Finally we end with conclusions in section \ref{conclusion}.

%======================================================================================================================
\section{Model of the compact object}\label{methodology}

It has already been shown that for an axially symmetric body, to emit gravitational wave, the rotation axis and the magnetic field axis should not be aligned to each other \citep{1996A&A...312..675B}. In other words, there must be a non-zero angle between the body's axis of symmetry and the rotation axis. The moment of inertia of a body is given by \citep{1996A&A...312..675B}
\begin{align*}
I_{jk} = \int \rho(x) \big(x_i x_i \delta_{jk}-x_j x_k\big) d^3x,
\end{align*}
where $\rho(x)$ is the density of the star at a distance $x$ from the center. For an axisymmetric star with $I_{xx}$, $I_{yy}$, $I_{zz}$ being the principal moments of inertia of the object about its three principal axes ($x-$, $y-$, $z-$axes) with $I_{xx}=I_{yy}$, situated at a distance $d$ from us, the dimensionless amplitudes of the two polarizations of the gravitational wave at time $t$, are given by \citep{1996A&A...312..675B, 1979PhRvD..20..351Z}
\begin{equation}\label{gravitational polarization}
\begin{aligned}
h_+ &= h_0\sin\chi\Bigg[\frac{1}{2}\cos i \sin i\cos\chi\cos\Omega t-\frac{1+\cos^2i}{2}\sin\chi\cos2\Omega t\Bigg],\\
h_\times &= h_0\sin\chi\Bigg[\frac{1}{2}\sin i\cos\chi\sin\Omega t-\cos i\sin\chi\sin2\Omega t\Bigg],
\end{aligned}
\end{equation}
with 
\begin{equation}\label{grav_wave_amplitude}
h_0 = -\frac{6G}{c^4}Q_{z'z'}\frac{\Omega^2}{d},
\end{equation}
where $Q_{z'z'}$ is the quadrupole moment of the distorted star, $\Omega$ the rotational frequency of the star, $\chi$ the angle between the rotation axis and the body's third principle axis $z$, $i$ the angle between the rotation axis of the object and our line of sight, $G$ Newton's gravitation constant and $c$ the speed of light. It is evident that if $\chi$ is small, the emission at frequency $\Omega$ is dominant. On the other hand, for large $\chi$, the emission at frequency $2\Omega$ will be prominent. In general, both the frequencies will be present in the radiation.

If the moments of inertia of an object about its three principle axes are known, then moment of inertia about any arbitrary axis $\vec{n}$ can be calculated as
\begin{align}
I_{nn}= \begin{pmatrix}
	\cos\alpha & \cos\beta & \cos\gamma
\end{pmatrix}
\begin{pmatrix}
	I_{xx} & I_{xy} & I_{xz}\\
	I_{xy} & I_{yy} & I_{yz}\\
	I_{xz} & I_{yz} & I_{zz}\\
\end{pmatrix}
\begin{pmatrix}
	\cos\alpha \\ \cos\beta \\ \cos\gamma
\end{pmatrix},
\end{align}
where $\cos\alpha$, $\cos\beta$, $\cos\gamma$ are the direction cosines of the arbitrary axis $\vec{n}$ and $I_{xy}$, $I_{xz}$, $I_{yz}$ are the products of inertia of the body. For an axially symmetric body, $I_{xy}= I_{xz}= I_{yz} = 0$ and $I_{xx} = I_{yy}$, which reduces the above formula to
\begin{align}\label{moment of inertia}
I_{nn} = I_{xx}\cos^2\alpha + I_{xx}\cos^2\beta + I_{zz}\cos^2\gamma.
\end{align}

\begin{figure}
\centering
\includegraphics[scale=0.50]{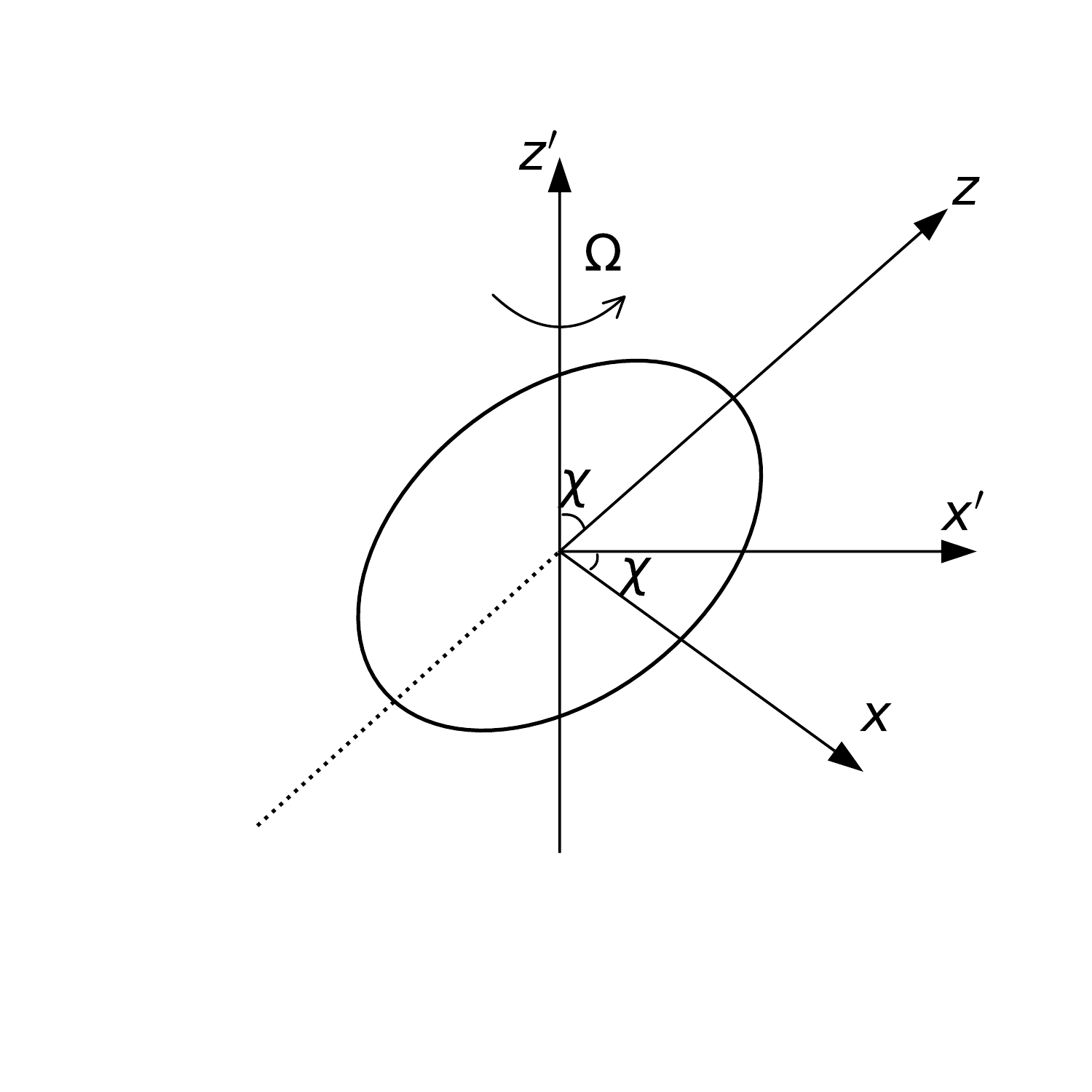}
\caption{A cartoon diagram of magnetized rotating white dwarf/neutron star with misalignment between magnetic field axis and rotation axis.}
\label{prolate}
\end{figure}

It was already shown that toroidal magnetic field makes a star prolate, whereas poloidal magnetic field as well as rotation deforms it to an oblate shape \citep{2002PhRvD..66h4025C,2004ApJ...600..296I, 2008PhRvD..78d4045K,2012MNRAS.427.3406F,2015JCAP...05..016D,2015MNRAS.454..752S,2015MNRAS.447.3475M,2016MNRAS.459.3407S}. In Figure \ref{prolate}, a cartoon diagram of white dwarf/neutron star is depicted such that the magnetic field is along $z$-axis and the star is rotating about the $z'$-axis which is at an angle $\chi$ with respect to z-axis. Hence, from the Figure \ref{prolate}, using equation (\ref{moment of inertia}), the moment of inertia about $x'-$, $y'-$, $z'-$axes are given by
\begin{equation}\label{new moment of inertia}
\begin{aligned}
I_{x'x'} &= I_{xx}\cos^2\chi+I_{zz}\sin^2\chi,\\
I_{y'y'} &= I_{yy},\\
I_{z'z'} &= I_{xx}\sin^2\chi+I_{zz}\cos^2\chi.
\end{aligned}
\end{equation}

Moreover, the relation between the quadrupole moment $Q_{ij}$ and moment of inertia $I_{ij}$ is given by
\begin{equation}
Q_{ij} = -I_{ij}+\frac{1}{3}I_{kk}\delta_{ij},
\end{equation}
where $\delta_{ij}$ is the Kronecker delta and $I_{kk} = I_{xx}+I_{yy}+I_{zz}$. Therefore, in primed frame, \begin{equation}\label{quadrupole moment}
Q_{i'j'} = -I_{i'j'}+\frac{1}{3}I_{k'k'}\delta_{i'j'}.
\end{equation}
Substituting this relation together with equations (\ref{new moment of inertia}) in equation (\ref{grav_wave_amplitude}), we have
\begin{equation}
h_0 = \frac{2G}{c^4}\frac{\Omega^2\epsilon I_{xx}}{d}(2\cos^2\chi-\sin^2\chi),
\end{equation}
where $\epsilon = (I_{zz}-I_{xx})/I_{xx}$ is the ellipticity of the body. The detailed derivation of this formula is explicitly shown in appendix \ref{appendix1}. As $\chi\to0$ (but $\neq 0$, otherwise no gravitational radiation), it reduces to
\begin{equation}\label{h0}
h_0 \to \frac{4G}{c^4}\frac{\Omega^2\epsilon I_{xx}}{d},
\end{equation}
which is exactly the same as given in equation (25) of Bonazzola and Gourgoulhon \citep{1996A&A...312..675B}.

To calculate the above-mentioned quantities such as $I_{xx}$, $I_{zz}$ etc., we use the {\it{XNS}} code, a numerical code to study the structure of neutron stars primarily\footnote{url: \url{http://www.arcetri.astro.it/science/ahead/XNS/code.html}}, but later appropriately modified for white dwarfs. This code solves for the axisymmetric equilibrium configuration of stellar structure in general Relativity. An axisymmetric equilibrium can be achieved, when the stars have rotation (uniform or differential) or magnetic field (toroidal or poloidal or mixed field) or both. In other words, the code solves the time independent general relativistic magnetohydrodynamic equations, thereby calculates the magnetostatic equilibrium. However, it is important to note that the star need not be in the ``stable'' equilibrium as per the {\it XNS} code. For example, the code generates the equilibrium structure of a star in the presence of very high magnetic field, although it is long known that the star might be unstable from the stability analysis, particularly for purely toroidal or poloidal field configurations. Nevertheless, the primary focus of the paper is on
the detectability of gravitational wave (GW) from white dwarfs and neutron stars with given
magnetic field strengths, but not geometries and the
explicit stability analysis of the configurations \citep{1973MNRAS.161..365T,1973MNRAS.162..339W,1973MNRAS.163...77M,1974MNRAS.168..505M,1985MNRAS.216..139P,2006A&A...453..687B,2007A&A...469..275B,2012MNRAS.424..482L}. The main choice that the code takes is the conformally flat metric in $3+1$ dimensions, which is given by \citep{2014MNRAS.439.3541P} $$ds^2= -\alpha^2dt^2+ \psi^4[dr^2 + r^2d\theta^2 + r^2\sin^2\theta(d\phi+\beta^\phi dt)^2],$$ with $(r,\theta, \phi)$ are the usual spherical polar coordinates, $\alpha(r,\theta)$ (also known as the lapse function) and $\psi(r,\theta)$ are the conformal factors, and $\beta^\phi$ is the shift vector in the $\phi$-direction. Moreover, the code implicitly assumes $\chi$ to be zero or does not include information about $\chi$ and, hence, we make small angle approximation to overcome this limitation, i.e. $\chi$ is close to zero, which implies that we can use equation (\ref{h0}) effectively. Since $\chi$ is small, radiation at the frequency $\Omega$ is dominant, as we have discussed above. Moreover, the amplitude of $h_+$ and $h_\times$ in equation (\ref{gravitational polarization}) will be suppressed by the other factors present therein. For instance, at $\chi=3\degree$,
\begin{align*}
\text{max}\Big\{\sin\chi\Big(\frac{1}{2}\cos i \sin i\cos\chi\cos\Omega t -\frac{1+\cos^2i}{2}\sin\chi\cos2\Omega t\Big)\Big\} \\= 0.0110297
\end{align*}
for $t=0$ and $i = i_{\max} \simeq 46.5\degree$. Hence maximum amplitude received by the detector at $\chi = 3\degree$ is $h = 0.0110297h_0$, which we consider for further calculations.

While estimating the structure of compact objects in the {\it{XNS}} code, we choose the number of grid points in radial and polar directions to be $N_r = 500$ and $N_\theta = 100$ respectively. The value of $R_\text{max}$ is chosen in such a way that it is always larger than the radius of the compact object. Moreover, since {\it{XNS}} code runs only when the equation of state is given in $P=K\rho^\Gamma$ form with $P$ being the pressure and $\rho$ the density, we choose $\Gamma = 4/3$ for high central density white dwarfs, $\rho_c \geq 10^8$ g cm$^{-3}$ with $K = (1/8)(3/\pi)^{1/3}hc/(\mu_e m_H)^{4/3}$, where $h$ is the Planck's constant, $\mu_e$ the mean molecular weight per electron and $m_H$ the mass of the hydrogen atom; whereas for low $\rho_c \leq 10^6$ g cm$^{-3}$, we consider $\gamma = 5/3$ and accordingly $K=(1/20)(3/\pi)^{2/3}h^2/(m_e (\mu_e m_H)^{5/3})$ with $m_e$ being the mass of an electron \citep{choudhuri_2010}. This choice is approximately true without compromising any major physics, as shown earlier \citep{2015JCAP...05..016D}, e.g. it reproduces Chandrasekhar limit. For in between $\rho_c$, there is no well-fitted polytropic index to be considered. We choose maximum $\rho_c = 2.2\times10^{10}$ g cm$^{-3}$ for white dwarfs, because above that $\rho_c$, the non-magnetized and non-rotating white dwarfs and the corresponding mass-radius relation, according to TOV equation solutions, become unstable. Important point to note that the effects of magnetic field and pycno-nuclear
reactions further may lead to the change of upper bound of $\rho_c$
\citep{2014PhRvC..89f5804V,2019ApJ...879...46O}, which we do not consider here. Further, we assume the distance between the white dwarf and the detector to be 100 pc ($1\text{pc} \approx 3.086\times10^{13}$ km) and that of neutron star and detector to be 10 kpc. Moreover, in case of neutron star, since we need a parametric equation of state, we use the parameters according to \cite{2014MNRAS.439.3541P}.

%======================================================================================================================
\section{Gravitational wave amplitude from various compact sources}\label{Result}

We consider purely toroidal and purely poloidal magnetic field cases separately for both B-WDs and B-NSs. Note however that, in reality, compact objects are expected to consist of mixed field geometry. Hence the actual results may be in between that of purely poloidal and purely toroidal cases \citep{1973MNRAS.161..365T,1973MNRAS.162..339W,1973MNRAS.163...77M,1974MNRAS.168..505M,1985MNRAS.216..139P,2006A&A...453..687B,2007A&A...469..275B,2012MNRAS.424..482L,2013PhRvL.110g1101L,2013MNRAS.435L..43C}. We explore the variation of $h_0$ with the change of different quantities such as central density, rotation, magnetic field etc. Moreover, we estimate $h_0$ if the white dwarfs are in a binary system. Subsequently, we display these estimated values of $h_0$ along with the sensitivity curves of different detectors.

\subsection{White dwarfs with purely toroidal magnetic field}
\begin{figure}
\centering
\subfigure[$\Omega \sim 0.0628$ rad s$^{-1}$, $B_\text{max} \sim 2.7\times 10^{14}$ G, ME/GE $\sim 0.1$, KE/GE $\sim 3.6\times10^{-6}$.]{\includegraphics[scale=0.5]{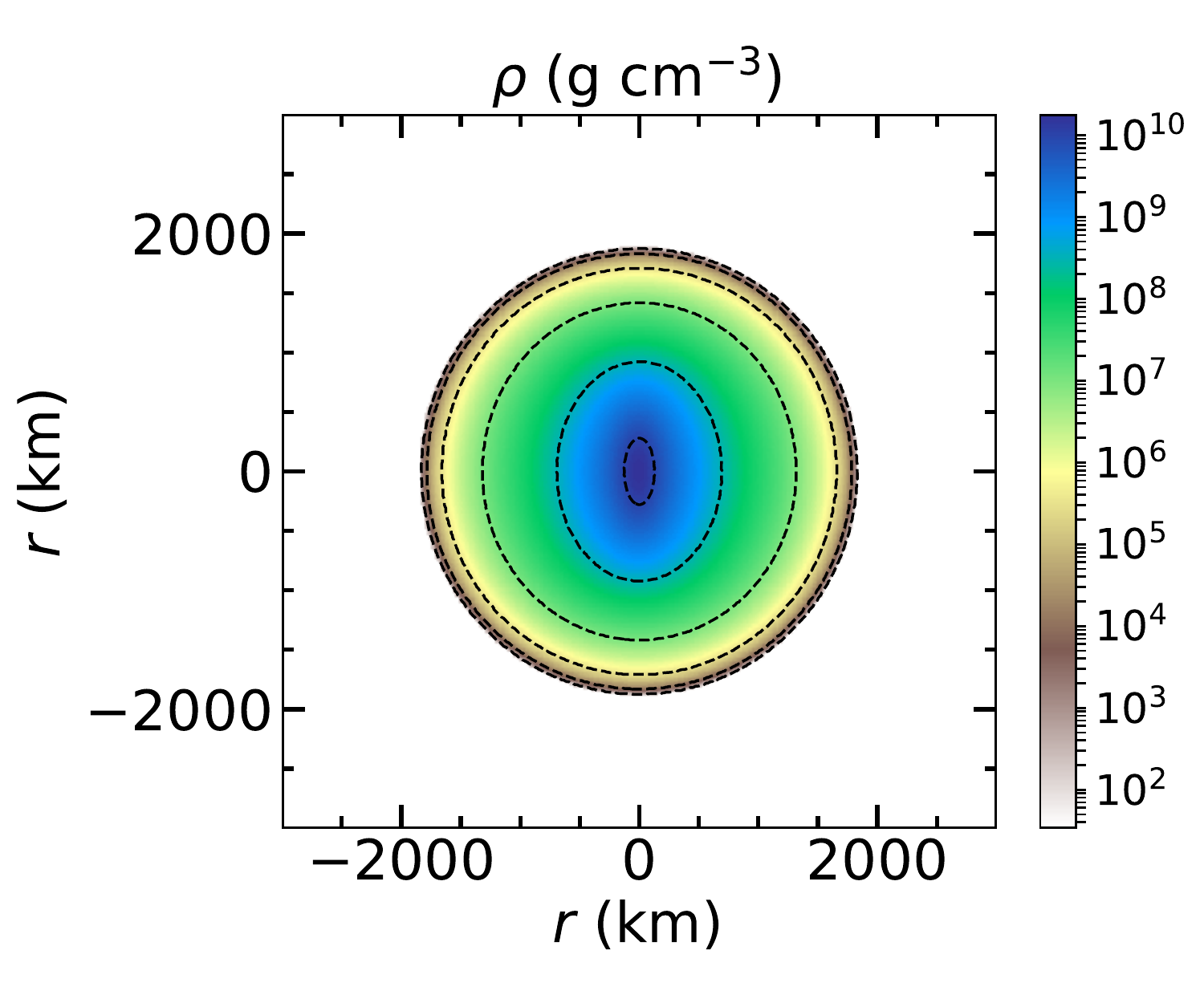}}
\subfigure[$\Omega \sim 3.6537$ rad s$^{-1}$, $B_\text{max} \sim 2.7\times 10^{14}$ G, ME/GE $\sim 0.1$, KE/GE $\sim 1.3\times10^{-2}$.]{\includegraphics[scale=0.5]{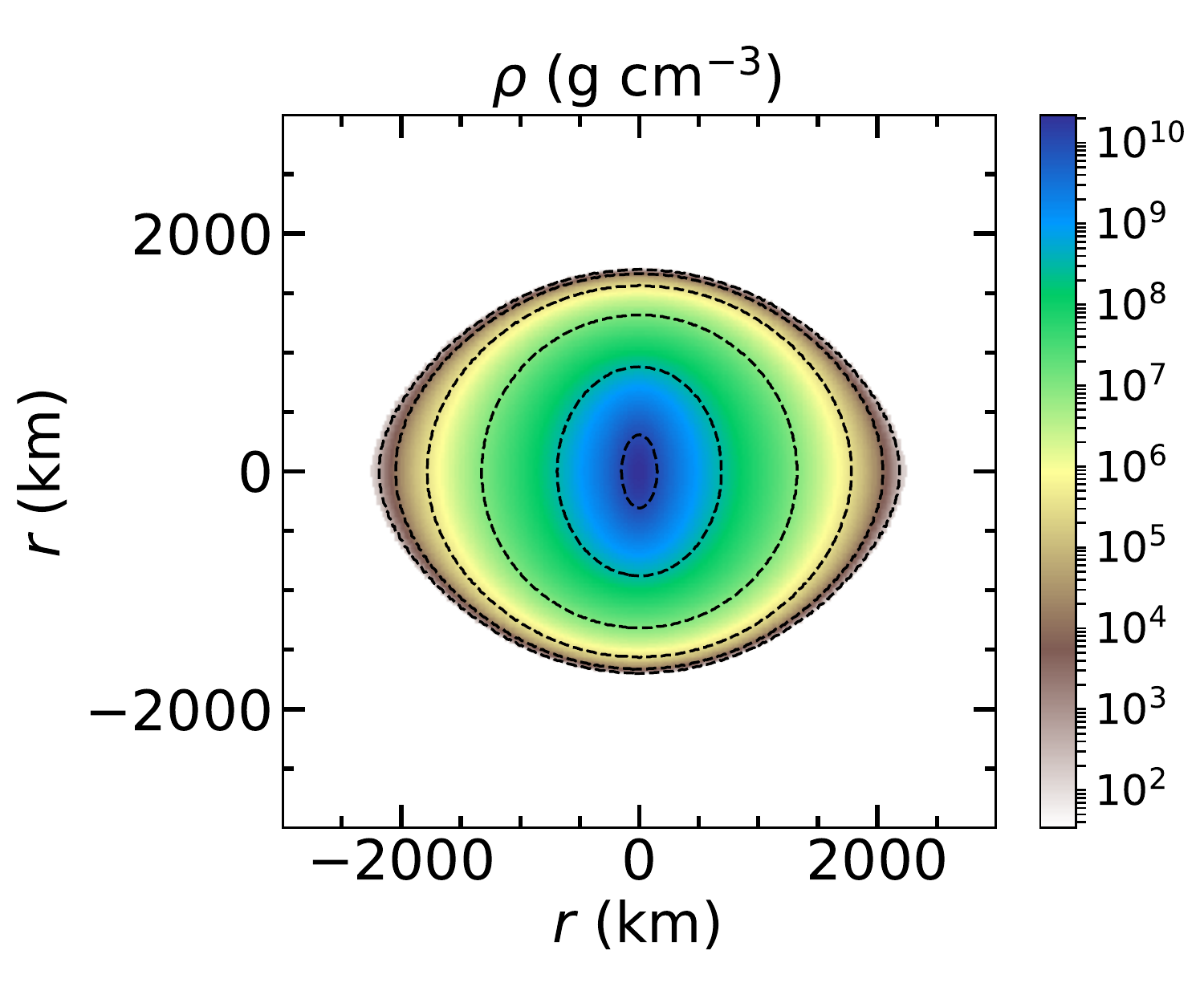}}
\caption{Density isocontours of uniformly rotating white dwarf with toroidal magnetic field.} \label{Tor_mag_field_figure}
\end{figure}

\begin{figure}
\centering
\includegraphics[scale=0.5]{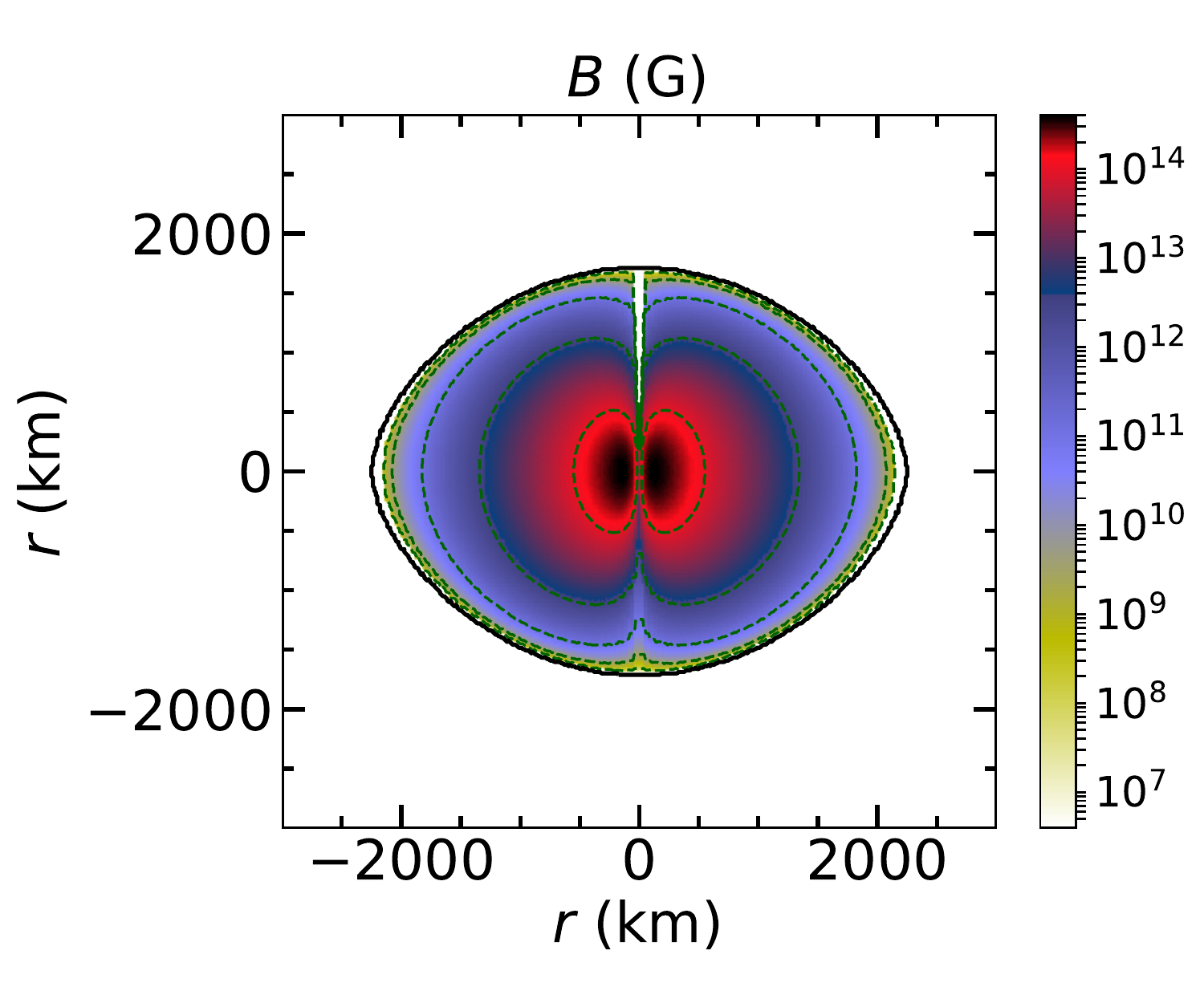}
\caption{Isocontours of magnetic field strength for uniformly rotating white dwarf with toroidal magnetic field. The black solid line represents the surface of the white dwarf. All the parameters are same as Figure \ref{Tor_mag_field_figure}(a).}
\label{Tor_mag_field_strength_figure}
\end{figure}

It was already shown that purely toroidal magnetic field not only makes the star prolate \citep{2002PhRvD..66h4025C,2004ApJ...600..296I, 2008PhRvD..78d4045K,2012MNRAS.427.3406F,2015JCAP...05..016D, 2015MNRAS.454..752S}, but also increases its equatorial radius. It is observed that the deformation at the core is more prominent than the outer region. Nevertheless, rotation of a star makes it oblate and hence there is always a competition between these two opposing effects to decide whether the star will be an overall oblate or prolate
(see also \citealt{2011MNRAS.417.2288M,2013MNRAS.434.1658M}). We show in Figure \ref{Tor_mag_field_figure} two typical cases for toroidal magnetic field configuration combined with the rotation. Figure \ref{Tor_mag_field_figure}(a) shows the density contour with the uniform angular frequency $\Omega = 0.0628$ rad s$^{-1}$ with kinetic to gravitational energy ratio, $\text{KE/GE}\sim 3.58\times10^{-6}$. Since, the angular frequency is small, it does not affect the star considerably, resulting in a prolate star. On the other hand, Figure \ref{Tor_mag_field_figure}(b) illustrates a star with angular frequency $\Omega = 3.6537$ rad s$^{-1}$ with $\text{KE/GE}\sim 1.33\times10^{-2}$, and due to this high angular velocity, the low density region is affected by the rotation more significantly than the high density region, resulting in an overall oblate shaped white dwarf. Here in both the cases, $\rho_c \sim 2.2\times 10^{10}$ g cm$^{-3}$, magnetic field at the center of the white dwarf $B_\text{max} \sim 2.7\times 10^{14}$ G. Indeed, a few white dwarfs are observed with the surface magnetic field $\sim10^9$ G \citep{2000MNRAS.317..310H,2005AJ....130..734V,2013ApJ...773...47B}, hence central field might be much larger than $10^9$ G. In fact, it has already been argued in the literature that the central field could be as large as $10^{14}-10^{15}$ G \citep{2015PhRvD..92h3006F,2017ApJ...843..131S}.
This is similar to the magnetic field in 
the case of the sun, where it is known from observations that 
the surface field is $\sim 1$ G. However, the central field could be
$\sim 10^4 - 10^7$ G, indicating an increase in $4-7$ orders of 
magnitude \citep{1979cmft.book.....P}.
Now, in time, sun-like stars will become white dwarfs 
and they may possess similar field profile as earlier, as argued by \cite{2014MPLA...2950035D}.
Hence, a white dwarf with $B_\text{max}$ considered here might have surface field $10^9$ G, which observed data have already confirmed. As a result, magnetic to gravitational energy ratio, $\text{ME/GE}\sim 0.1$. The surface of a white dwarf is determined, when the density decreases at least up to $7-8$ orders in magnitude compared to its central density, i.e. $\rho_c \sim 10^{7-8} \rho_\text{surf}$ (ideally zero), where $\rho_\text{surf}$ is the density at the surface. The bar-code shows the density of the white dwarf at different radii. The typical isocontours of magnetic field strength are shown in Figure \ref{Tor_mag_field_strength_figure}. It is confirmed herein that the surface magnetic field can decrease up to $\sim 10^9$ G even if the central field $\sim 10^{14}$ G. However, GW astronomy may also help in identifying magnetized white dwarfs with surface fields higher than $10^9$ G, as argued below also. 

\begin{table*}
\centering
\caption{Uniformly rotating white dwarf with toroidal magnetic field ($d = 100$ pc) with $\chi = 3\degree$. Here $B_\text{max}$ is the maximum magnetic field close to the center of white dwarf, when surface field could be much smaller.}
\label{Toroidal Magnetic Field Table}
\begin{tabular}{|l|l|l|l|l|l|l|l|l|l|}
\hline
$\rho_c$ (g cm$^{-3}$) & $M$ ($M_\odot$)& $R_E$ (km) & $R_P/R_E$ & $B_{\max}$ (G) & $\nu$ (Hz) & ME/GE & KE/GE & $|I_{x'x'}-I_{y'y'}|/I_{z'z'}$ & $h_0$\\
\hline\hline
	$2.2\times 10^{10}$ & 1.405 & 1179.9 & 1.0000 & $7.693\times 10^{13}$ & 0.0100 & $4.78\times10^{-3}$ & $1.85\times 10^{-6}$ & $3.38\times10^{-5}$ & $1.5560\times 10^{-25}$\\
	& 1.405 & 1179.9 & 1.0000 & $7.693\times 10^{13}$ & 0.0500 & $4.79\times10^{-3}$ & $4.63\times 10^{-5}$ & $3.35\times10^{-5}$ & $3.8539\times 10^{-24}$\\
	& 1.406 & 1179.9 & 1.0000 & $7.694\times 10^{13}$ & 0.1000 & $4.79\times10^{-3}$ & $1.85\times 10^{-4}$ & $3.25\times10^{-5}$ & $1.4964\times 10^{-23}$\\
	& 1.426 & 1330.5 & 0.8868 & $7.720\times 10^{13}$ & 0.7107 & $4.82\times10^{-3}$ & $9.80\times 10^{-3}$ & $3.77\times10^{-5}$ & $9.3190\times 10^{-22}$\\
\hline
	$10^{10}$ & 1.418 & 1531.3 & 1.0000 & $4.692\times 10^{13}$ & 0.0100 & $5.06\times10^{-3}$ & $4.04\times 10^{-6}$ & $3.60\times10^{-5}$ & $2.8447\times 10^{-25}$\\
	& 1.417 & 1531.3 & 1.0000 & $4.575\times 10^{13}$ & 0.0500 & $4.80\times10^{-3}$ & $1.01\times 10^{-4}$ & $3.35\times10^{-5}$ & $6.6038\times 10^{-24}$\\
	& 1.418 & 1531.3 & 1.0000 & $4.575\times 10^{13}$ & 0.1000 & $4.80\times10^{-3}$ & $4.04\times 10^{-3}$ & $3.13\times10^{-5}$ & $2.4731\times 10^{-23}$\\
	& 1.433 & 1681.9 & 0.9104 & $4.586\times 10^{13}$ & 0.4200 & $4.83\times10^{-3}$ & $7.36\times 10^{-3}$ & $1.92\times10^{-5}$ & $2.7886\times 10^{-22}$\\
\hline
	$10^9$ & 1.435 & 3338.7 & 1.0150 & $9.923\times 10^{12}$ & 0.0050 & $4.83\times10^{-3}$ & $9.98\times 10^{-6}$ & $3.47\times10^{-5}$ & $3.2414\times 10^{-25}$\\
	& 1.435 & 3338.7 & 1.0150 & $9.923\times 10^{12}$ & 0.0100 & $4.83\times10^{-3}$ & $3.99\times 10^{-5}$ & $3.45\times10^{-5}$ & $1.2887\times 10^{-24}$\\
	& 1.437 & 3388.9 & 0.9852 & $9.927\times 10^{12}$ & 0.0500 & $4.83\times10^{-3}$ & $1.00\times 10^{-3}$ & $2.76\times10^{-5}$ & $2.5885\times 10^{-23}$\\
	& 1.458 & 3790.6 & 0.8676 & $9.967\times 10^{12}$ & 0.1615 & $4.87\times10^{-3}$ & $1.10\times 10^{-2}$ & $4.47\times10^{-5}$ & $4.6693\times 10^{-22}$\\
\hline
	$10^8$ & 1.441 & 7254.8 & 1.0069 & $2.147\times 10^{12}$ & 0.0010 & $4.84\times10^{-3}$ & $3.98\times 10^{-6}$ & $3.47\times10^{-5}$ & $6.0755\times 10^{-26}$\\
	& 1.441 & 7307.2 & 1.0000 & $2.147\times 10^{12}$ & 0.0050 & $4.84\times10^{-3}$ & $9.97\times 10^{-5}$ & $3.40\times10^{-5}$ & $1.4896\times 10^{-24}$\\
	& 1.441 & 7305.0 & 1.0000 & $2.147\times 10^{12}$ & 0.0100 & $4.84\times10^{-3}$ & $3.99\times 10^{-4}$ & $3.19\times10^{-5}$ & $5.5918\times 10^{-24}$\\
	& 1.464 & 8158.6 & 0.8769 & $2.155\times 10^{12}$ & 0.0500 & $4.88\times10^{-3}$ & $1.05\times 10^{-2}$ & $4.08\times10^{-5}$ & $1.9042\times 10^{-22}$\\
\hline\hline
	$2.2\times10^{10}$ & 1.677 & 1705.5 & 1.0260 & $2.729\times 10^{14}$ & 0.0100 & 0.1003 & $3.58\times 10^{-6}$ & $7.63\times10^{-4}$ & $6.8065\times 10^{-24}$\\
	& 1.677 & 1705.5 & 1.0260 & $2.729\times 10^{14}$ & 0.0500 & 0.1003 & $8.95\times 10^{-5}$ & $7.62\times10^{-4}$ & $1.7002\times 10^{-22}$\\
	& 1.677 & 1714.4 & 1.0207 & $2.729\times 10^{14}$ & 0.1000 & 0.1003 & $3.59\times 10^{-4}$ & $7.57\times10^{-4}$ & $6.7828\times 10^{-22}$\\
	& 1.696 & 1962.5 & 0.8736 & $2.719\times 10^{14}$ & 0.5000 & 0.0990 & $9.52\times 10^{-3}$ & $6.16\times10^{-4}$ & $1.4766\times 10^{-20}$\\
	& 1.706 & 2237.1 & 0.7624 & $2.719\times 10^{14}$ & 0.5815 & 0.0991 & $1.33\times 10^{-2}$ & $5.58\times10^{-4}$ & $1.8788\times 10^{-20}$\\
	& 1.938 & 2804.0 & 0.7757 & $3.106\times 10^{14}$ & 0.4361 & 0.1406 & $1.01\times 10^{-2}$ & $1.04\times10^{-3}$ & $3.1995\times 10^{-20}$\\
	& 1.940 & 2981.4 & 0.7296 & $3.106\times 10^{14}$ & 0.4458 & 0.1405 & $1.07\times 10^{-2}$ & $1.03\times10^{-3}$ & $3.3189\times 10^{-20}$\\

\hline
	$10^{10}$ & 1.684 & 2219.4 & 1.0240 & $1.617\times 10^{14}$ & 0.0100 & 0.1004 & $7.87\times 10^{-6}$ & $7.67\times10^{-4}$ & $1.1686\times 10^{-23}$\\
	& 1.685 & 2219.4 & 1.0240 & $1.617\times 10^{14}$ & 0.0500 & 0.1005 & $1.97\times 10^{-4}$ & $7.65\times10^{-4}$ & $2.9163\times 10^{-22}$\\
	& 1.686 & 2237.1 & 1.0158 & $1.617\times 10^{14}$ & 0.1000 & 0.1005 & $7.91\times 10^{-4}$ & $7.56\times10^{-4}$ & $1.1597\times 10^{-21}$\\
	& 1.701 & 2458.6 & 0.9135 & $1.614\times 10^{14}$ & 0.3000 & 0.0999 & $7.45\times 10^{-3}$ & $6.56\times10^{-4}$ & $9.5406\times 10^{-21}$\\
	& 1.945 & 3707.9 & 0.7587 & $1.837\times 10^{14}$ & 0.3000 & 0.1400 & $1.06\times 10^{-2}$ & $1.03\times10^{-3}$ & $2.5607\times 10^{-20}$\\
	& 1.946 & 3831.9 & 0.7341 & $1.838\times 10^{14}$ & 0.3037 & 0.1400 & $1.08\times 10^{-2}$ & $1.02\times10^{-3}$ & $2.6061\times 10^{-20}$\\

\hline
	$10^9$ & 1.692 & 4785.8 & 1.0216 & $3.474\times 10^{13}$ & 0.0050 & 0.0992 & $1.95\times 10^{-5}$ & $7.60\times10^{-4}$ & $1.3549\times 10^{-23}$\\
	& 1.692 & 4785.9 & 1.0216 & $3.474\times 10^{13}$ & 0.0100 & 0.0992 & $7.80\times 10^{-5}$ & $7.60\times10^{-4}$ & $5.4164\times 10^{-23}$\\
	& 1.697 & 4902.5 & 0.9952 & $3.473\times 10^{13}$ & 0.0500 & 0.0992 & $1.98\times 10^{-3}$ & $7.34\times10^{-4}$ & $1.3285\times 10^{-21}$\\
	& 1.713 & 5398.7 & 0.8950 & $3.474\times 10^{13}$ & 0.1000 & 0.0994 & $8.30\times 10^{-3}$ & $6.44\times10^{-4}$ & $4.9346\times 10^{-21}$\\
	& 1.960 & 8443.6 & 0.7314 & $3.966\times 10^{13}$ & 0.0937 & 0.1392 & $1.03\times 10^{-2}$ & $1.05\times10^{-3}$ & $1.1956\times 10^{-20}$\\
\hline
	$10^8$ & 1.698 & 10330.7 & 1.0240 & $7.491\times 10^{12}$ & 0.0010 & 0.0993 & $7.79\times 10^{-6}$ & $7.63\times10^{-4}$ & $2.5485\times 10^{-24}$\\
	& 1.712 & 10507.9 & 1.0202 & $7.579\times 10^{12}$ & 0.0050 & 0.1034 & $2.01\times 10^{-4}$ & $7.94\times10^{-4}$ & $6.8323\times 10^{-23}$\\
	& 1.700 & 10401.6 & 1.0170 & $7.491\times 10^{12}$ & 0.0100 & 0.0993 & $7.83\times 10^{-4}$ & $7.53\times10^{-4}$ & $2.5292\times 10^{-22}$\\
	& 1.717 & 11429.4 & 0.9121 & $7.493\times 10^{12}$ & 0.0300 & 0.0996 & $7.42\times 10^{-3}$ & $6.59\times10^{-4}$ & $2.1126\times 10^{-21}$\\
\hline
\end{tabular}
\end{table*}

Table \ref{Toroidal Magnetic Field Table} shows different $h_0$ for various $\rho_c$. In the table, $M$ represents the mass of compact object, $R_E$ the equatorial radius, $R_P$ the polar radius and $\nu$ the linear frequency defined as $\nu=\Omega/2\pi$. It is observed that for a given $\rho_c$, $h_0$ increases with the increase in rotation because $h_0\propto \Omega^2$. Moreover, if we compare two different central density cases with rotation being fixed, it is observed that $h_0$ is larger for smaller $\rho_c$ white dwarf. This is because the radius of the star is larger for lower central density and hence moment of inertia increases. Since $h_0 \propto \epsilon I_{xx}$, the value of $h_0$ increases for smaller $\rho_c$ white dwarf. However, it may not be true always, as the smaller white dwarfs can rotate much faster and the above-mentioned argument is true only if the white dwarfs have the same angular frequency. Therefore, combining 
the effects of rotation and size (and density) of the white dwarf optimally, $h_0$ is calculated. For all the cases, ME/GE as well as KE/GE are chosen to be $\lesssim0.1$ to maintain stable equilibrium \citep{1953ApJ...118..116C,1989MNRAS.237..355K, 2009MNRAS.397..763B}. It is noticed that $h_0$ increases as ME/GE increases because as the magnetic field strength in the star increases, it deviates more from spherical geometry, resulting in higher quadrupole moment. From the table, it is also clear that highly magnetized white dwarfs are indeed super-Chandrasekhar candidates.

%-------------------------------------------------------------------------------------------------------------------

\subsection{White dwarfs with purely poloidal magnetic field}

\begin{figure}
\centering
\includegraphics[scale=0.5]{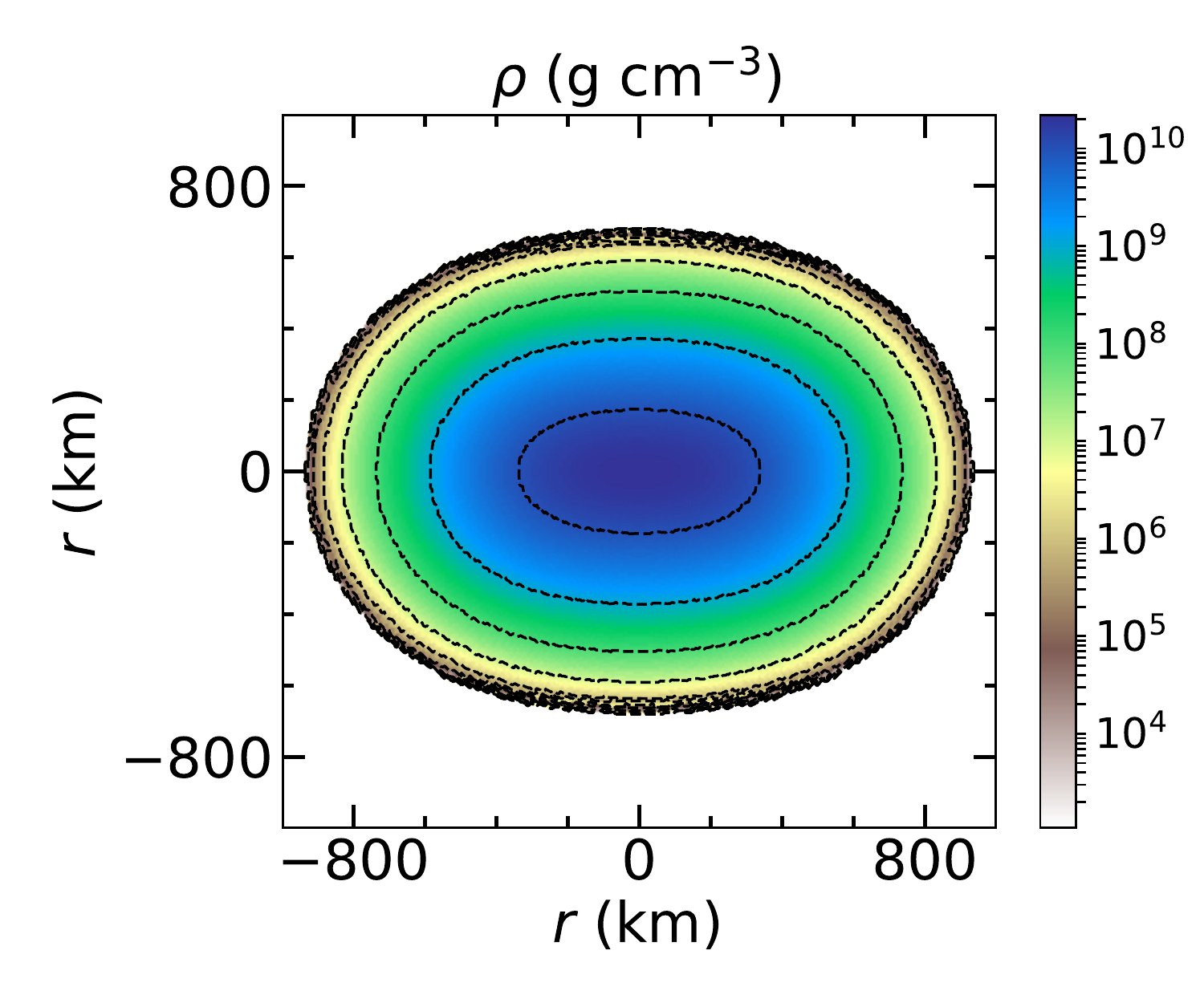}
\caption{Density isocontours of uniformly rotating white dwarf with poloidal magnetic field. Here $\Omega \sim 2.0297$ rad s$^{-1}$, $B_\text{max} \sim 5.3\times 10^{14}$ G, ME/GE $\sim 0.1$, KE/GE $\sim 2.5\times10^{-3}$.}
\label{Pol_mag_field_figure}
\end{figure}

\begin{figure}
\centering
\includegraphics[scale=0.5]{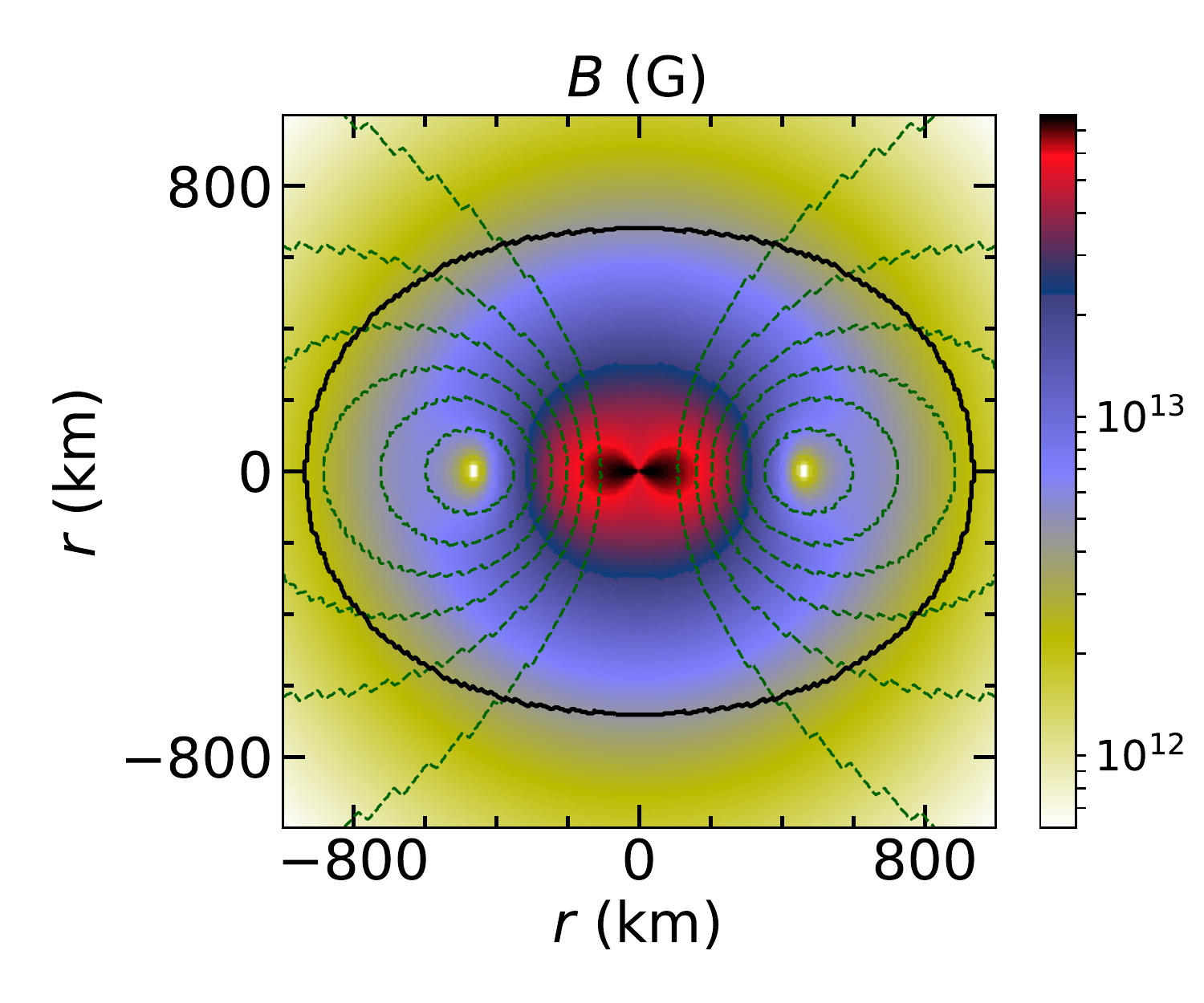}
\caption{Same as Figure \ref{Tor_mag_field_strength_figure}, except for poloidal magnetic field. All the parameters are same as Figure \ref{Pol_mag_field_figure}.}
\label{Pol_mag_field_strength_figure}
\end{figure}

A similar exploration is carried out for uniformly rotating white dwarfs with purely poloidal magnetic field. It was already discussed \citep{2004ApJ...600..296I, 2015JCAP...05..016D, 2015MNRAS.454..752S} that purely poloidal magnetic field as well as rotation both make the star oblate. Hence in order to obtain stable maximally deformed white dwarfs, resulting in from both these effects, we have to precisely adjust the value of magnetic field and rotation. Figure \ref{Pol_mag_field_figure} illustrates a typical case for this configuration. Here $\rho_c \sim 2\times10^{10}$ g cm$^{-3}$, $\Omega \sim 2.03$ rad s$^{-1}$, $B_\text{max} \sim 5.29\times 10^{14}$ G, $\text{ME/GE} \sim 0.1$ and $\text{KE/GE} \sim 2.488\times 10^{-3}$. Table \ref{Poloidal Magnetic Field Table} shows different values of $h_0$ with the change of central density and angular frequency. The isocontours of poloidal magnetic field strength are shown in Figure \ref{Pol_mag_field_strength_figure}. It is evident that the surface field need not be very low if the central field is high enough, unlike for the case of the toroidal magnetic field, according to {\it XNS} code. In reality, since the stable white dwarfs are expected to possess a mixture of toroidal and poloidal magnetic fields, depending on their relative strengths, the surface field can have smaller to larger values for strong central magnetic field.

\begin{table*}
\centering
\caption{Uniformly rotating white dwarf with poloidal magnetic field ($d = 100$ pc) with $\chi = 3\degree$. Here $B_\text{max}$ is the maximum magnetic field at the center of white dwarf, when surface field could be much smaller.}
\label{Poloidal Magnetic Field Table}
\begin{tabular}{|l|l|l|l|l|l|l|l|l|l|}
\hline
$\rho_c$ (g cm$^{-3}$) & $M$ ($M_\odot$)& $R_E$ (km) & $R_P/R_E$ & $B_{\max}$ (G) & $\nu$ (Hz) & ME/GE & KE/GE & $|I_{x'x'}-I_{y'y'}|/I_{z'z'}$ & $h_0$\\
\hline\hline
	$2.2\times 10^{10}$ & 1.405 & 1156.2 & 0.9770 & $1.004\times 10^{14}$ & 0.0100 & $4.74\times 10^{-3}$ & $4.04\times 10^{-5}$ & $6.36\times10^{-5}$ & $2.7087\times 10^{-25}$\\
	& 1.405 & 1156.2 & 0.9770 & $1.004\times 10^{14}$ & 0.0500 & $4.75\times 10^{-3}$ & $2.36\times 10^{-4}$ & $6.39\times10^{-5}$ & $6.8023\times 10^{-24}$\\
	& 1.406 & 1156.2 & 0.9770 & $1.004\times 10^{14}$ & 0.1000 & $4.75\times 10^{-3}$ & $5.56\times 10^{-4}$ & $6.48\times10^{-5}$ & $2.7595\times 10^{-23}$\\
	& 1.408 & 1200.5 & 0.9336 & $1.009\times 10^{14}$ & 0.5000 & $4.85\times 10^{-3}$ & $6.28\times 10^{-3}$ & $9.27\times10^{-5}$ & $1.0124\times 10^{-21}$\\
	& 1.410 & 1306.8 & 0.8508 & $9.824\times 10^{13}$ & 0.8076 & $4.67\times 10^{-3}$ & $1.49\times 10^{-2}$ & $1.38\times10^{-4}$ & $4.1220\times 10^{-21}$\\
	& 1.410 & 1333.4 & 0.8538 & $1.504\times 10^{13}$ & 0.8076 & $1.08\times 10^{-4}$ & $1.47\times 10^{-2}$ & $8.35\times10^{-5}$ & $2.4628\times 10^{-21}$\\
	& 1.420 & 1342.3 & 0.8482 & $3.342\times 10^{11}$ & 0.8173 & $5.33\times 10^{-8}$ & $1.50\times 10^{-2}$ & $8.43\times10^{-5}$ & $2.5502\times 10^{-21}$\\
\hline
	$10^{10}$ & 1.418 & 1510.6 & 0.9824 & $5.912\times 10^{13}$ & 0.0100 & $4.76\times 10^{-3}$ & $6.98\times 10^{-5}$ & $6.39\times10^{-5}$ & $4.6798\times 10^{-25}$\\
	& 1.418 & 1510.6 & 0.9824 & $5.912\times 10^{13}$ & 0.0500 & $4.77\times 10^{-3}$ & $4.24\times 10^{-4}$ & $6.45\times10^{-5}$ & $1.1817\times 10^{-23}$\\
	& 1.418 & 1510.6 & 0.9765 & $5.915\times 10^{13}$ & 0.1000 & $4.78\times 10^{-3}$ & $1.04\times 10^{-3}$ & $6.64\times10^{-5}$ & $4.8744\times 10^{-23}$\\
	& 1.420 & 1661.2 & 0.8720 & $5.980\times 10^{13}$ & 0.5000 & $4.99\times 10^{-3}$ & $1.33\times 10^{-2}$ & $1.30\times10^{-4}$ & $2.5286\times 10^{-21}$\\
	& 1.425 & 1758.7 & 0.8237 & $5.736\times 10^{13}$ & 0.5815 & $4.63\times 10^{-3}$ & $1.76\times 10^{-2}$ & $1.50\times10^{-4}$ & $4.0397\times 10^{-21}$\\
	& 1.432 & 1971.3 & 0.7438 & $1.296\times 10^{13}$ & 0.6461 & $2.30\times 10^{-4}$ & $2.10\times 10^{-2}$ & $1.22\times10^{-4}$ & $4.0757\times 10^{-21}$\\
\hline
	$10^9$ & 1.415 & 3233.9 & 0.9836 & $1.324\times 10^{13}$ & 0.0100 & $5.06\times 10^{-3}$ & $3.43\times 10^{-4}$ & $6.78\times10^{-5}$ & $2.3221\times 10^{-24}$\\
	& 1.415 & 3269.3 & 0.9729 & $1.264\times 10^{13}$ & 0.0500 & $4.63\times 10^{-3}$ & $2.47\times 10^{-3}$ & $6.79\times10^{-5}$ & $5.8348\times 10^{-23}$\\
	& 1.415 & 3340.2 & 0.9469 & $1.268\times 10^{13}$ & 0.1000 & $4.70\times 10^{-3}$ & $6.90\times 10^{-3}$ & $8.70\times10^{-5}$ & $3.0428\times 10^{-22}$\\
	& 1.425 & 3783.2 & 0.8173 & $1.283\times 10^{13}$ & 0.1874 & $4.93\times 10^{-3}$ & $2.01\times 10^{-2}$ & $1.55\times10^{-4}$ & $2.0313\times 10^{-21}$\\
	& 1.446 & 4474.3 & 0.7069 & $5.510\times 10^{12}$ & 0.2099 & $1.05\times 10^{-3}$ & $5.90\times 10^{-2}$ & $1.37\times10^{-4}$ & $2.3507\times 10^{-21}$\\
\hline
	$10^8$ & 1.413 & 6999.4 & 0.9797 & $2.797\times 10^{12}$ & 0.0010 & $4.88\times 10^{-3}$ & $1.46\times 10^{-4}$ & $6.48\times10^{-5}$ & $1.0375\times 10^{-25}$\\
	& 1.413 & 6999.4 & 0.9797 & $2.798\times 10^{12}$ & 0.0050 & $4.89\times 10^{-3}$ & $8.08\times 10^{-4}$ & $6.54\times10^{-5}$ & $2.6189\times 10^{-24}$\\
	& 1.414 & 6999.4 & 0.9797 & $2.799\times 10^{12}$ & 0.0100 & $4.91\times 10^{-3}$ & $1.81\times 10^{-3}$ & $6.73\times10^{-5}$ & $1.0793\times 10^{-23}$\\
	& 1.430 & 7672.7 & 0.8799 & $2.712\times 10^{12}$ & 0.0500 & $4.75\times 10^{-3}$ & $1.74\times 10^{-2}$ & $1.25\times10^{-4}$ & $5.2936\times 10^{-22}$\\
	& 1.452 & 9444.7 & 0.7036 & $2.740\times 10^{12}$ & 0.0678 & $4.96\times 10^{-3}$ & $2.96\times 10^{-4}$ & $1.87\times10^{-4}$ & $1.5440\times 10^{-21}$\\
\hline
	$10^6$ & 0.473 & 11020.3 & 0.9818 & $1.078\times 10^{11}$ & 0.0010 & $4.89\times 10^{-3}$ & $3.47\times 10^{-3}$ & $6.15\times10^{-5}$ & $2.2462\times 10^{-25}$\\
	& 0.476 & 11121.4 & 0.9684 & $1.050\times 10^{11}$ & 0.0050 & $4.84\times 10^{-3}$ & $2.26\times 10^{-2}$ & $8.22\times10^{-5}$ & $7.6372\times 10^{-24}$\\
	& 0.486 & 11422.0 & 0.9297 & $1.037\times 10^{11}$ & 0.0100 & $5.02\times 10^{-3}$ & $6.06\times 10^{-2}$ & $1.55\times10^{-4}$ & $6.1053\times 10^{-23}$\\
	& 0.510 & 12275.5 & 0.8405 & $9.071\times 10^{10}$ & 0.0162 & $4.24\times 10^{-3}$ & $1.39\times 10^{-1}$ & $3.06\times10^{-4}$ & $3.6118\times 10^{-22}$\\
\hline\hline
	$2.2\times 10^{10}$ & 1.615 & 961.3 & 0.7512 & $4.830\times 10^{14}$ & 0.0100 & 0.0977 & $7.42\times 10^{-4}$ & $7.55\times10^{-4}$ & $4.0205\times 10^{-24}$\\
	& 1.615 & 961.3 & 0.7512 & $4.830\times 10^{14}$ & 0.0500 & 0.0978 & $4.03\times 10^{-4}$ & $7.55\times10^{-4}$ & $1.0055\times 10^{-22}$\\
	& 1.616 & 961.3 & 0.7512 & $4.832\times 10^{14}$ & 0.1000 & 0.0980 & $8.85\times 10^{-4}$ & $7.55\times10^{-4}$ & $4.0269\times 10^{-22}$\\
	& 1.631 & 970.2 & 0.7352 & $4.875\times 10^{14}$ & 0.5000 & 0.1006 & $7.68\times 10^{-3}$ & $7.70\times10^{-4}$ & $1.0472\times 10^{-20}$\\
	& 1.639 & 987.9 & 0.7220 & $4.839\times 10^{14}$ & 0.7107 & 0.1001 & $1.35\times 10^{-2}$ & $7.75\times10^{-4}$ & $2.1634\times 10^{-20}$\\
	& 1.699 & 908.1 & 0.7073 & $5.645\times 10^{14}$ & 0.0032 & 0.1111 & $1.01\times 10^{-5}$ & $8.57\times10^{-4}$ & $4.9764\times 10^{-25}$\\
\hline
	$10^{10}$ & 1.627 & 1244.8 & 0.7509 & $2.897\times 10^{14}$ & 0.0100 & 0.0998 & $1.26\times 10^{-4}$ & $7.63\times10^{-4}$ & $6.9822\times 10^{-24}$\\
	& 1.628 & 1244.8 & 0.7509 & $2.897\times 10^{14}$ & 0.0500 & 0.1000 & $7.02\times 10^{-4}$ & $7.64\times10^{-4}$ & $1.7470\times 10^{-22}$\\
	& 1.628 & 1253.7 & 0.7456 & $2.893\times 10^{14}$ & 0.1000 & 0.1000 & $1.58\times 10^{-3}$ & $7.63\times10^{-4}$ & $6.9880\times 10^{-22}$\\
	& 1.640 & 1298.0 & 0.7270 & $2.829\times 10^{14}$ & 0.5000 & 0.0983 & $1.54\times 10^{-2}$ & $7.68\times10^{-4}$ & $1.8144\times 10^{-20}$\\
	& 1.702 & 1182.8 & 0.7154 & $3.325\times 10^{14}$ & 0.0032 & 0.1100 & $1.73\times 10^{-5}$ & $8.54\times10^{-4}$ & $8.4732\times 10^{-25}$\\
\hline
	$10^9$ & 1.622 & 2697.9 & 0.7548 & $6.131\times 10^{13}$ & 0.0050 & 0.0968 & $1.53\times 10^{-4}$ & $7.52\times10^{-4}$ & $8.0470\times 10^{-24}$\\
	& 1.622 & 2697.9 & 0.7548 & $6.131\times 10^{13}$ & 0.0100 & 0.0969 & $3.23\times 10^{-4}$ & $7.52\times10^{-4}$ & $3.2197\times 10^{-23}$\\
	& 1.625 & 2718.5 & 0.7490 & $6.143\times 10^{13}$ & 0.0500 & 0.0976 & $2.33\times 10^{-3}$ & $7.55\times10^{-4}$ & $8.1205\times 10^{-22}$\\
	& 1.636 & 2739.2 & 0.7358 & $6.182\times 10^{13}$ & 0.1000 & 0.0992 & $6.46\times 10^{-3}$ & $7.65\times10^{-4}$ & $3.3388\times 10^{-21}$\\
	& 1.635 & 2821.9 & 0.7289 & $5.972\times 10^{13}$ & 0.1615 & 0.0946 & $1.43\times 10^{-2}$ & $7.56\times10^{-4}$ & $8.7428\times 10^{-21}$\\
\hline
	$10^8$ & 1.623 & 5865.3 & 0.7583 & $1.312\times 10^{13}$ & 0.0010 & 0.0960 & $1.38\times 10^{-4}$ & $7.47\times10^{-4}$ & $1.4946\times 10^{-24}$\\
	& 1.623 & 5865.3 & 0.7583 & $1.312\times 10^{13}$ & 0.0050 & 0.0962 & $7.63\times 10^{-4}$ & $7.47\times10^{-4}$ & $3.7399\times 10^{-23}$\\
	& 1.624 & 5865.3 & 0.7523 & $1.313\times 10^{13}$ & 0.0100 & 0.0965 & $1.71\times 10^{-3}$ & $7.48\times10^{-4}$ & $1.5001\times 10^{-22}$\\
	& 1.659 & 5971.6 & 0.7211 & $1.340\times 10^{13}$ & 0.0500 & 0.1020 & $1.60\times 10^{-2}$ & $7.81\times10^{-4}$ & $4.0977\times 10^{-21}$\\
\hline
	$10^6$ & 0.610 & 11723.2 & 0.7259 & $5.057\times 10^{11}$ & 0.0010 & 0.0995 & $3.58\times 10^{-3}$ & $7.40\times10^{-4}$ & $4.6022\times 10^{-24}$\\
	& 0.609 & 11773.4 & 0.7271 & $4.950\times 10^{11}$ & 0.0050 & 0.1000 & $2.39\times 10^{-2}$ & $7.35\times10^{-4}$ & $1.1435\times 10^{-22}$\\
	& 0.616 & 12024.5 & 0.7119 & $4.788\times 10^{11}$ & 0.0100 & 0.1000 & $6.54\times 10^{-2}$ & $7.47\times10^{-4}$ & $4.8253\times 10^{-22}$\\
	& 0.640 & 12626.9 & 0.6700 & $4.581\times 10^{11}$ & 0.0155 & 0.1001 & $1.40\times 10^{-1}$ & $7.91\times10^{-4}$ & $1.3601\times 10^{-21}$\\
\hline
\end{tabular}
\end{table*}

%-------------------------------------------------------------------------------------------------------------------

\subsection{Magnetized white dwarfs with differential rotation}
\begin{figure}
\centering
\subfigure[Toroidal magnetic field with $\Omega \sim 0.62$ rad s$^{-1}$ at surface, 
$\Omega_c \sim 10.15$ rad s$^{-1}$, $B_\text{max} \sim 3.2\times 10^{14}$ G, ME/GE $ \sim 0.14$, KE/GE $\sim 0.1$.]{\includegraphics[scale=0.5]{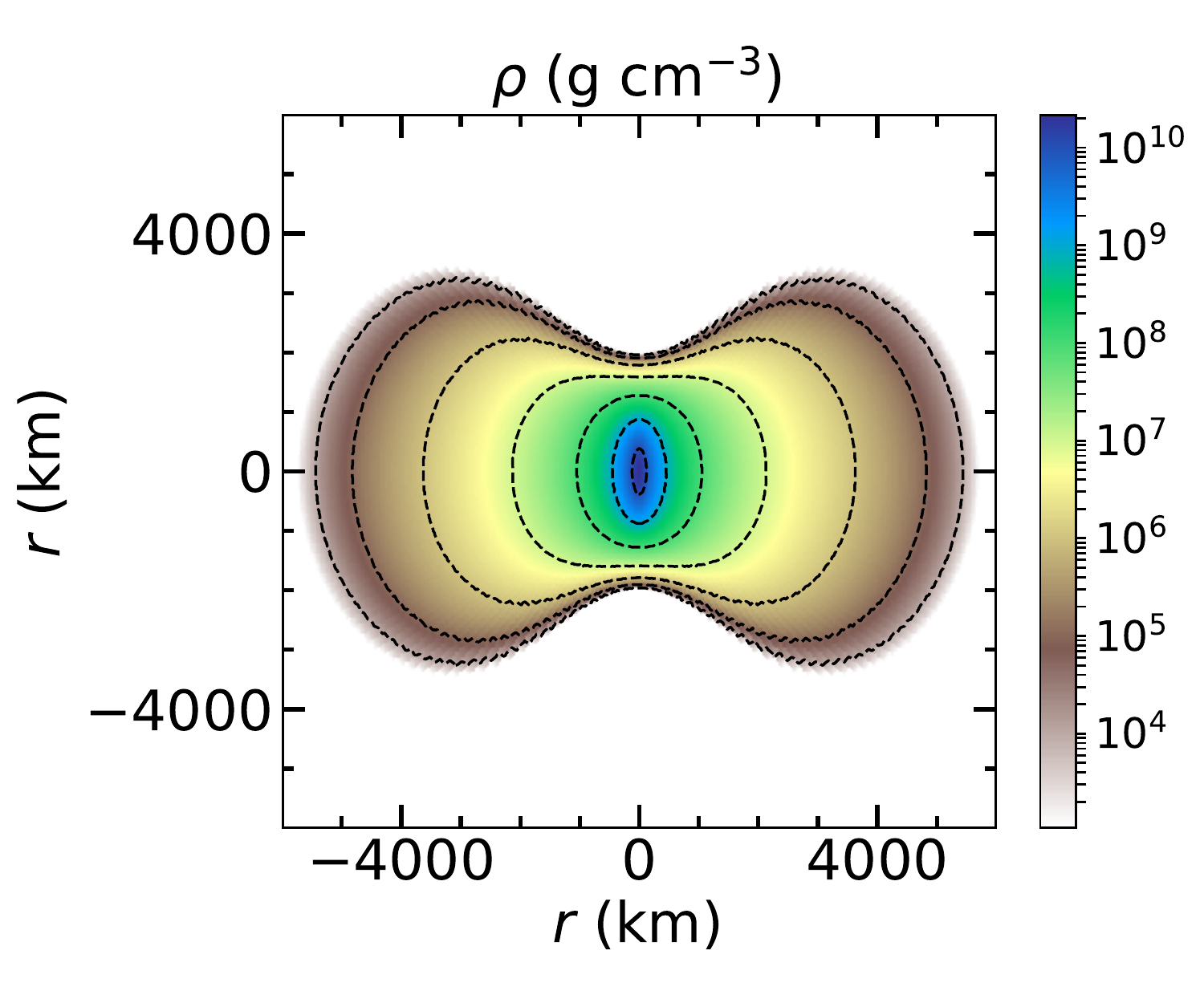}}
\subfigure[Poloidal magnetic field with $\Omega \sim 2.74$ rad s$^{-1}$ at surface, 
$\Omega_c \sim 20.30$ rad s$^{-1}$, $B_\text{max} \sim 3.9\times 10^{11}$ G, ME/GE $\sim 7.8\times10^{-8}$, KE/GE $\sim 0.14$.]{\includegraphics[scale=0.5]{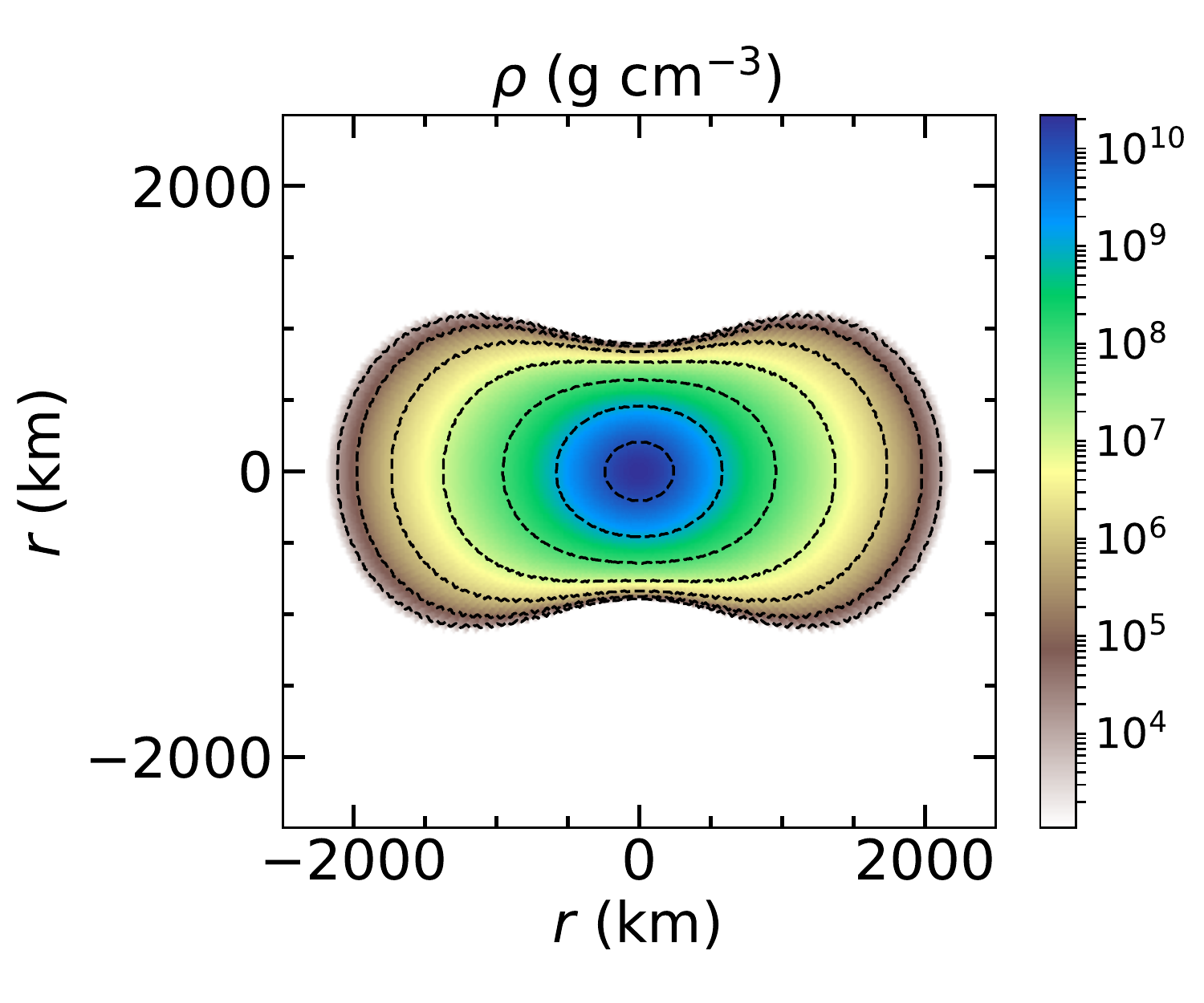}}
\caption{Density isocontours of differentially rotating magnetized white dwarf.}
\label{Differential rotation}
\end{figure}

Although it is not known whether a white dwarf possesses and/or sustains differential rotation or not, we explore $h_0$ considering differential rotation too. A detailed discussion about differentially rotating white dwarfs is given by \cite{2015MNRAS.454..752S}. We assume differential rotation together with purely toroidal and poloidal magnetic field cases separately. 
The angular velocity profile in {\it XNS} code is given by 
(\citealt{2003LRR.....6....3S,2011A&A...528A.101B})
%\cite{2003LRR.....6....3S,2011A&A...528A.101B}
\begin{equation}
F(\Omega) = A^2 (\Omega_c - \Omega)=
\frac{R^2\left(\Omega-\omega\right)}{\alpha^2-R^2\left(\Omega-\omega\right)^2}, 
\end{equation}
where $A$ is a constant indicating the extent of differential rotation, 
$R=\psi^2 r\sin\theta$, $\omega=-\beta^\phi$,
$\Omega_c$ the angular velocity at the center and $\Omega$ is that at $r$
for an observer at infinity (also known as the zero angular momentum observer: ZAMO).
If $A\to\infty$, then $\Omega_c = \Omega$ which implies uniform rotation.
From Figure \ref{Differential rotation}, it is well noted that `polar hollow' structure can be formed in the cases of differential rotation regardless of the geometry of the magnetic field. Tables \ref{Differntial Toroidal Magnetic Field Table} and \ref{Differntial Poloidal Magnetic Field Table} show different $h_0$ for differentially rotating white dwarfs with purely toroidal and purely poloidal magnetic fields respectively for $\rho_c = 2.2\times 10^{10}$ g cm$^{-3}$. 

\begin{table*}
\centering
\caption{Differentially rotating white dwarf with toroidal magnetic field, $\rho_c = 2.2\times10^{10}$ g cm$^{-3}$. Here $\Omega = \Omega_s$, the angular frequency at the surface and $d = 100$ pc with $\chi = 3\degree$. $B_\text{max}$ is the maximum magnetic field close to the center of white dwarf, when surface field could be much smaller.}
\label{Differntial Toroidal Magnetic Field Table}
\begin{tabular}{|l|l|l|l|l|l|l|l|l|l|l|}
\hline
$M$ ($M_\odot$)& $R_E$ (km) & $R_P/R_E$ & $B_{\max}$ (G) & $\Omega_c$ (rad s$^{-1}$) & $\Omega$ (rad s$^{-1}$) & $\nu$ (Hz) & ME/GE & KE/GE & $|I_{x'x'}-I_{y'y'}|/I_{z'z'}$ & $h_0$\\
\hline\hline
1.451 & 1227.1 & 0.9350 & $7.456\times10^{12}$ & 10.1486 & 1.2774 & 0.2033 & $4.40\times10^{-3}$ & 0.0152 & $4.95\times10^{-5}$ & $1.0023\times10^{-22}$ \\
1.520 & 1785.3 & 0.6179 & $7.495\times10^{12}$ & 10.1486 & 4.1100 & 0.6541 & $4.46\times10^{-3}$ & 0.0449 & $2.62\times10^{-4}$ & $6.8748\times10^{-21}$ \\
1.531 & 4651.5 & 0.2343 & $7.490\times10^{12}$ & 10.1486 & 1.2526 & 0.1994 & $4.46\times10^{-3}$ & 0.0498 & $3.31\times10^{-4}$ & $8.6992\times10^{-22}$ \\
1.556 & 1302.4 & 0.7823 & $7.759\times10^{12}$ & 20.2972 & 2.3001 & 0.3661 & $4.72\times10^{-3}$ & 0.0618 & $2.57\times10^{-4}$ & $2.0436\times10^{-21}$ \\
1754 & 1918.2 & 0.4919 & $7.894\times10^{12}$ & 20.2972 & 3.0509 & 0.4856 & $4.96\times10^{-3}$ & 0.1349 & $5.88\times10^{-4}$ & $1.3417\times10^{-20}$ \\
1.800 & 996.7 & 0.5911 & $1.135\times10^{13}$ & 121.7831 & 2.3435 & 0.3730 & $4.70\times10^{-3}$ & 0.1520 & $4.68\times10^{-4}$ & $3.4198\times10^{-21}$ \\
2.621 & 3583.8 & 0.4759 & $3.166\times10^{14}$ & 20.2972 & 0.7952 & 0.1266 & 0.1396 & 0.1402 & $5.00\times10^{-6}$ & $4.2652\times10^{-23}$ \\
2.902 & 8693.1 & 0.2342 & $3.204\times10^{14}$ & 10.1486 & 0.3391 & 0.0540 & 0.1409 & 0.1410 & $6.42\times10^{-4}$ & $4.3313\times10^{-21}$ \\
\hline

\end{tabular}
\end{table*}

\begin{table*}
\centering
\caption{Differentially rotating white dwarf with poloidal magnetic field, $\rho_c = 2.2\times10^{10}$ g cm$^{-3}$. Here $\Omega = \Omega_s$, the angular frequency at the surface and $d = 100$ pc with $\chi = 3\degree$. $B_\text{max}$ is the maximum magnetic field at the center of white dwarf, when surface field could be much smaller.}
\label{Differntial Poloidal Magnetic Field Table}
\begin{tabular}{|l|l|l|l|l|l|l|l|l|l|l|}
\hline
$M$ ($M_\odot$)& $R_E$ (km) & $R_P/R_E$ & $B_{\max}$ (G) & $\Omega_c$ (rad s$^{-1}$) & $\Omega$ (rad s$^{-1}$) & $\nu$ (Hz) & ME/GE & KE/GE & $|I_{x'x'}-I_{y'y'}|/I_{z'z'}$ & $h_0$\\
\hline\hline
1.400 & 1191.7 & 0.9405 & $3.368\times10^{11}$ & 10.1486 & 1.3445 & 0.2140 & $5.35\times10^{-8}$ & 0.0186 & $7.74\times10^{-5}$ & $1.5831\times10^{-22}$ \\
1.447 & 1687.8 & 0.6378 & $3.458\times10^{11}$ & 10.1486 & 4.4132 & 0.7024 & $5.81\times10^{-8}$ & 0.0465 & $2.68\times10^{-4}$ & $7.2002\times10^{-21}$ \\
1.456 & 3758.1 & 0.2849 & $3.461\times10^{11}$ & 10.1486 & 1.9603 & 0.3120 & $5.89\times10^{-8}$ & 0.0503 & $3.15\times10^{-4}$ & $1.7571\times10^{-21}$ \\
1.487 & 1271.4 & 0.7840 & $3.589\times10^{11}$ & 20.2972 & 2.4302 & 0.3868 & $5.99\times10^{-8}$ & 0.0647 & $2.79\times10^{-4}$ & $2.2389\times10^{-21}$ \\
1.717 & 2246.0 & 0.4043 & $3.909\times10^{11}$ & 20.2972 & 2.7416 & 0.4363 & $7.81\times10^{-8}$ & 0.1403 & $6.78\times10^{-4}$ & $1.3059\times10^{-20}$ \\
1.466 & 1111.9 & 0.8088 & $3.704\times10^{11}$ & 60.8916 & 1.0505 & 0.1672 & $5.41\times10^{-8}$ & 0.0558 & $1.98\times10^{-4}$ & $2.4860\times10^{-22}$ \\
1.613 & 1111.9 & 0.6892 & $4.102\times10^{11}$ & 60.8916 & 2.0631 & 0.3284 & $6.07\times10^{-8}$ & 0.1107 & $4.01\times10^{-4}$ & $2.3098\times10^{-21}$ \\
1.708 & 1120.8 & 0.6364 & $4.340\times10^{11}$ & 60.8916 & 2.6425 & 0.4206 & $6.52\times10^{-8}$ & 0.1401 & $5.02\times10^{-4}$ & $5.3216\times10^{-21}$ \\
1.398 & 1182.8 & 0.9176 & $1.025\times10^{14}$ & 10.1486 & 1.3619 & 0.2168 & $5.05\times10^{-3}$ & 0.0189 & $1.39\times10^{-4}$ & $2.9575\times10^{-22}$ \\
1.460 & 1616.9 & 0.6493 & $1.054\times10^{14}$ & 10.1486 & 4.6012 & 0.7323 & $5.64\times10^{-3}$ & 0.0486 & $3.19\times10^{-4}$ & $9.3751\times10^{-21}$ \\
1.476 & 1103.1 & 0.7992 & $1.051\times10^{14}$ & 60.8916 & 1.0671 & 0.1698 & $4.59\times10^{-3}$ & 0.0586 & $2.45\times10^{-4}$ & $3.2178\times10^{-22}$ \\
1.654 & 1103.1 & 0.6707 & $1.058\times10^{14}$ & 60.8916 & 2.2963 & 0.3655 & $4.49\times10^{-3}$ & 0.1362 & $4.68\times10^{-4}$ & $3.4935\times10^{-21}$ \\
1.674 & 856.5 & 0.7241 & $5.448\times10^{14}$ & 20.2972 & 0.5830 & 0.0928 & $1.11\times10^{-1}$ & 0.0064 & $8.51\times10^{-4}$ & $3.9607\times10^{-22}$ \\
\hline
\end{tabular}
\end{table*}

In the Gaia-DR2 catalogue \citep{2018MNRAS.480.4505J}, 73,221 white dwarfs were reported. The maximum mass of this catalog is $1.4567M_\odot$ and minimum is $0.208M_\odot$. Moreover, the reported range of radius is $1474.5-17262$ km. Furthermore, \cite{2006Natur.443..308H}, \cite{2010ApJ...713.1073S} etc. argued the progenitor mass of overluminous type Ia supernovae could be as high high $\sim2.8M_\odot$. Hence, the masses reported by us in Tables \ref{Toroidal Magnetic Field Table} - \ref{Differntial Poloidal Magnetic Field Table} are quite in accordance with observation. While most of the corresponding radii also obey observation, smaller radius white dwarfs may indeed be so dim to be detected by the current technique, as argued by \cite{2018MNRAS.477.2705B}. They are also in accordance with other recent theoretical predictions, in particularly in the presence of magnetic fields (e.g. \citealt{2019ApJ...879...46O}).

All the values of $h_0$ presented in Tables \ref{Toroidal Magnetic Field Table} - \ref{Differntial Poloidal Magnetic Field Table} are displayed in Figures \ref{Detector} and \ref{Detector2} along with the various sensitivity curves of different detectors\footnote{\url{http://gwplotter.com/} and \url{http://www.srl.caltech.edu/~shane/sensitivity/}} (\citealt{2009LRR....12....2S,2015CQGra..32a5014M}, and the references therein). From the figures, it is well understood that larger the angle $\chi$ be, stronger is the gravitational radiation. It is also noticed that many of them can easily be detected by DECIGO, BBO and ALIA, whereas hardly any of them can be detected by LISA and eLISA directly. However, it is noticed that the highly magnetized white dwarfs can be detected by 1 year integration curve of LISA. Since the equatorial radius of the white dwarf increases with rotation, there is very little possibility of having any point above 1Hz frequency and hence it is hard to detect them by Einstein telescope.

%---------------------------------------------------------------------------------------------------------------------

\subsection{White dwarfs in a binary system}
We explore the strength of the gravitational radiation if the white dwarfs including B-WDs have a binary companion. For simplicity, we assume that this binary companion is also another white dwarf. For such a system, the dimensionless amplitude of the polarization is given by \citep{2006MNRAS.371.1231R, 1997CQGra..14.1525J}
\begin{eqnarray}
h &=& 2.84\times10^{-22}\sqrt{\cos^4i+6\cos^2i+1} \\ \nonumber &\times & \Bigg(\frac{M_c}{M_\odot}\Bigg)^{5/3}\Bigg(\frac{P_{orb}}{1 \text{hr}}\Bigg)^{-2/3}\Bigg(\frac{d}{1 \text{kpc}}\Bigg)^{-1},
\end{eqnarray}
where $P_{orb}$ is the orbital period of the binary system and $M_c$ is the chirp mass which is defined as 
\begin{equation}
M_c= \Bigg[\frac{m_1^3m_2^3}{(m_1+m_2)}\Bigg]^{1/5},
\end{equation}
where $m_1$ and $m_2$ are masses of the component white dwarfs. Moreover, the frequency of the gravitational wave ($f$) is double the orbital frequency ($f_{orb}$), i.e. $f=2f_{orb}$. In our calculation, we choose $d = 100$ pc and $i=0\degree$.

\begin{figure*}
\centering
\includegraphics[scale=0.5]{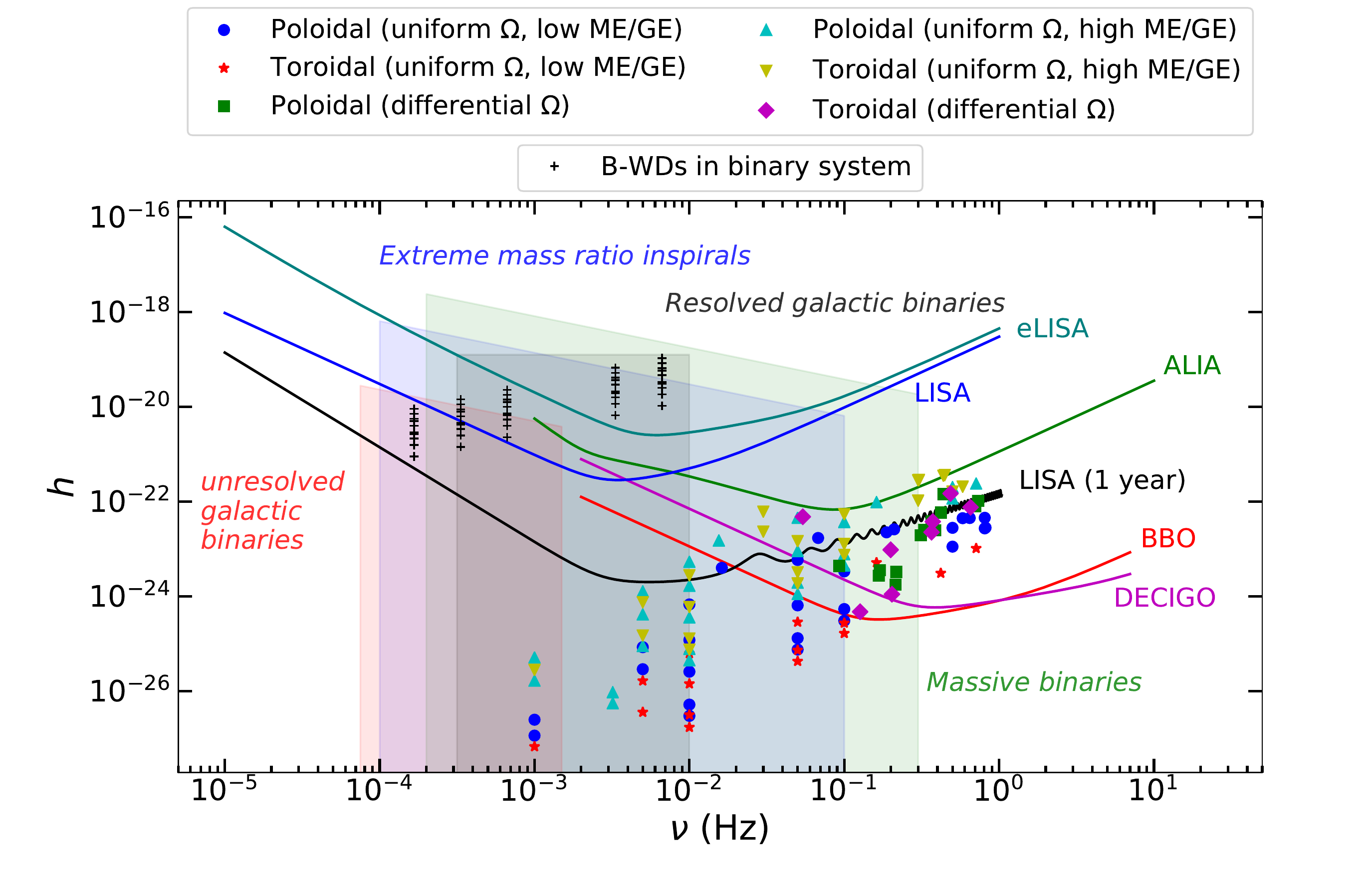}
\caption{Dimensionless gravitational wave amplitude for white dwarfs as a function of frequency, as given in Tables \ref{Toroidal Magnetic Field Table} - \ref{Differntial Poloidal Magnetic Field Table}, along with the sensitivity curves of various detectors. Here $h = 0.0110297h_0$ with $\chi = 3\degree$.}
\label{Detector}
\end{figure*}

\begin{figure*}
\centering
\includegraphics[scale=0.5]{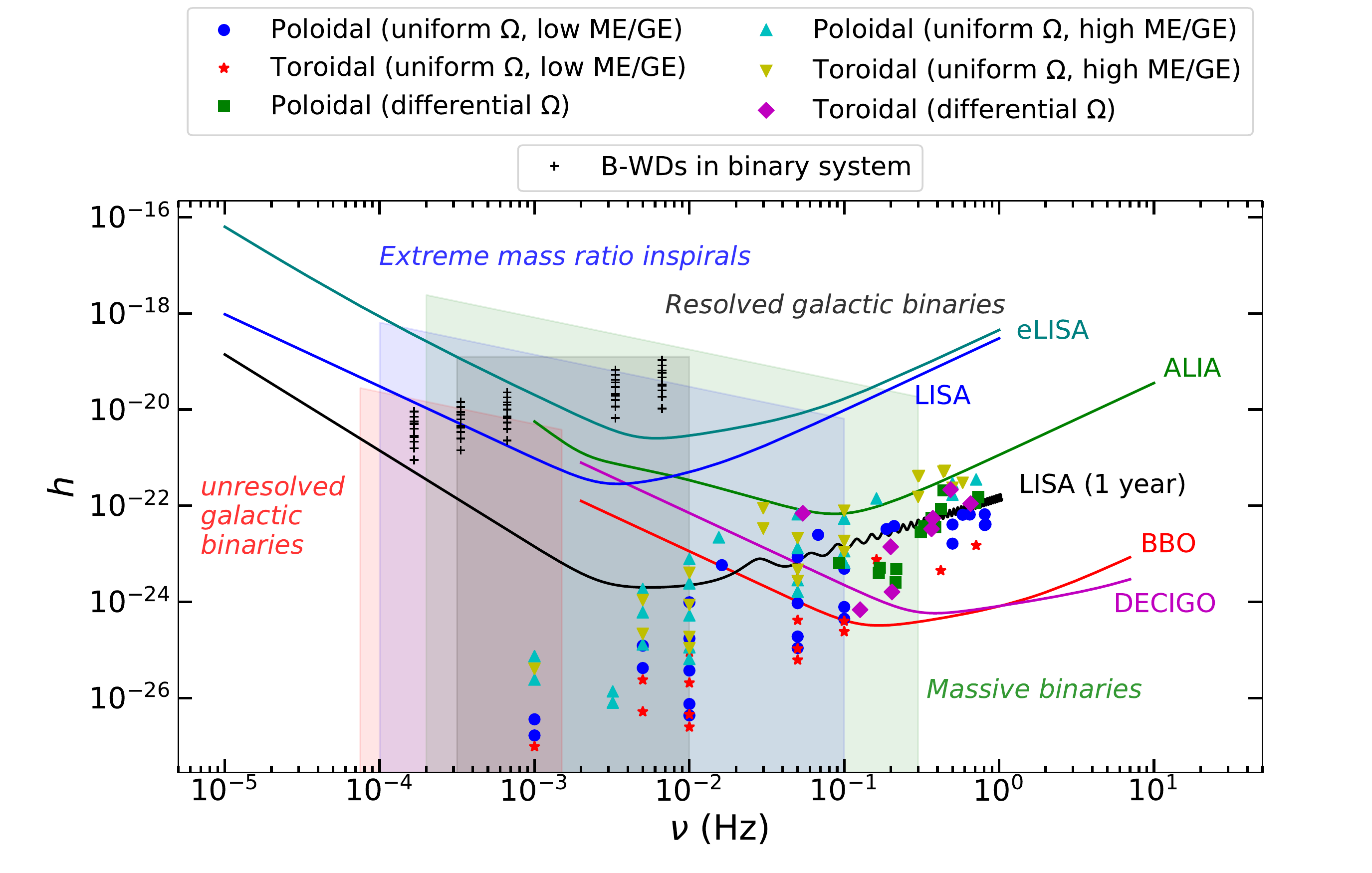}
\caption{Same as Figure \ref{Detector} except $h = 0.016098h_0$ with $\chi = 5\degree$.}
\label{Detector2}
\end{figure*}

We assume various orbital periods for different combinations of the white dwarfs with different masses based on previous literature \citep{2010ApJ...711L.138R, 2000MNRAS.317..310H, 2011ApJ...737L..23B,2018MNRAS.480..302K}. We choose the masses of the white dwarfs including B-WDs in the range $0.5-2 M_\odot$ (when the masses in a binary need not be the same) and vary orbital period from $5-200$ mins. The values of $h$ for these combinations are given in Table \ref{Table: Binary WD} and also shown in Figures \ref{Detector} and \ref{Detector2}. From this figure, it is clear that the strength of gravitational radiation from these binary systems is much higher than the strength of isolated rotating B-WDs. However, it is also evident from the figure that the frequency range for these systems is different from that of rotating B-WDs. Hence the detection of isolated rotating B-WDs and white dwarfs in binary systems, are clearly distinguishable from each other, if their respective distances are known. Indeed, distances of many white dwarfs are known independently \citep{1994PASP..106..209P,1999PASP..111..702A,2000MNRAS.317..310H}. In fact, we propose that based on $h$ and $\nu$ at which it is detected, B-WDs can be identified in GW astronomy. For example, if $h$ for a source is detected by 1 year integration curve of LISA, but not by LISA or eLISA directly, the source could be a B-WD. It is also noted from Figures \ref{Detector} and \ref{Detector2} that some of these B-WDs are clearly distinguished from the other galactic and extra-galactic sources and thereby avoiding any sorts of confusion noise \citep{2010ApJ...717.1006R,2015CQGra..32a5014M,2019CQGra..36j5011R}. In fact, confusion noise may further go away, once we concentrate on the integrated effect of the source of interest when combined effects from other sources may be canceled out. Moreover, with the help of proper source modeling, this problem of confusion noise can be negotiated, as people have done it for the EMRI sources \citep{2017PhRvD..95j3012B,van_de_Meent_2017}. However our present work is beyond the scope of source modeling.

\begin{table*}
\centering
\caption{Gravitational wave from white dwarf binaries with $d=100$ pc. }
\label{Table: Binary WD}
\begin{tabular}{|l|l|l|l|l|l|l|}
\hline
$m_1$ ($M_\odot$)& $m_2$ ($M_\odot$) & & & $h$ & &\\
\hline
& & $P_{orb}=5$ min & $P_{orb}=10$ min & $P_{orb}=50$ min & $P_{orb}=100$ min & $P_{orb}=200$ min \\ 
\hline\hline
0.5 & 0.5 & $3.72\times10^{-21}$ & $2.34\times10^{-21}$ & $8.02\times10^{-22}$ & $5.05\times10^{-22}$ & $3.18\times10^{-22}$\\
0.5 & 1.0 & $6.50\times10^{-21}$ & $4.10\times10^{-21}$ & $1.40\times10^{-21}$ & $8.82\times10^{-22}$ & $5.56\times10^{-22}$\\
0.5 & 1.5 & $8.86\times10^{-21}$ & $5.58\times10^{-21}$ & $1.91\times10^{-21}$ & $1.20\times10^{-21}$ & $7.58\times10^{-22}$\\
0.5 & 2.0 & $1.10\times10^{-20}$ & $6.91\times10^{-21}$ & $2.36\times10^{-21}$ & $1.49\times10^{-21}$ & $9.38\times10^{-22}$\\
1.0 & 1.0 & $1.18\times10^{-20}$ & $7.44\times10^{-21}$ & $2.55\times10^{-21}$ & $1.60\times10^{-21}$ & $1.01\times10^{-21}$\\
1.0 & 1.5 & $1.65\times10^{-20}$ & $1.04\times10^{-20}$ & $3.54\times10^{-21}$ & $2.23\times10^{-21}$ & $1.41\times10^{-21}$\\
1.0 & 2.0 & $2.06\times10^{-20}$ & $1.30\times10^{-20}$ & $4.45\times10^{-21}$ & $2.80\times10^{-21}$ & $1.76\times10^{-21}$\\
1.5 & 1.5 & $2.32\times10^{-20}$ & $1.46\times10^{-20}$ & $5.00\times10^{-21}$ & $3.15\times10^{-21}$ & $1.99\times10^{-21}$\\
1.5 & 2.0 & $2.94\times10^{-20}$ & $1.85\times10^{-20}$ & $6.34\times10^{-21}$ & $3.99\times10^{-21}$ & $2.51\times10^{-21}$\\
2.0 & 2.0 & $3.75\times10^{-20}$ & $2.36\times10^{-20}$ & $8.08\times10^{-21}$ & $5.09\times10^{-21}$ & $3.21\times10^{-21}$\\
\hline
\end{tabular}
\end{table*}

%---------------------------------------------------------------------------------------------------------------------
\subsection{Magnetized uniformly rotating neutron stars}
\begin{figure}
\centering
\subfigure[Toroidal magnetic field with $\Omega \sim 314.2$ rad s$^{-1}$, $B_\text{max} \sim 8.4\times 10^{16}$ G, ME/GE $\sim 3.7\times10^{-2}$, KE/GE $\sim 2.1\times10^{-3}$.]{\includegraphics[scale=0.5]{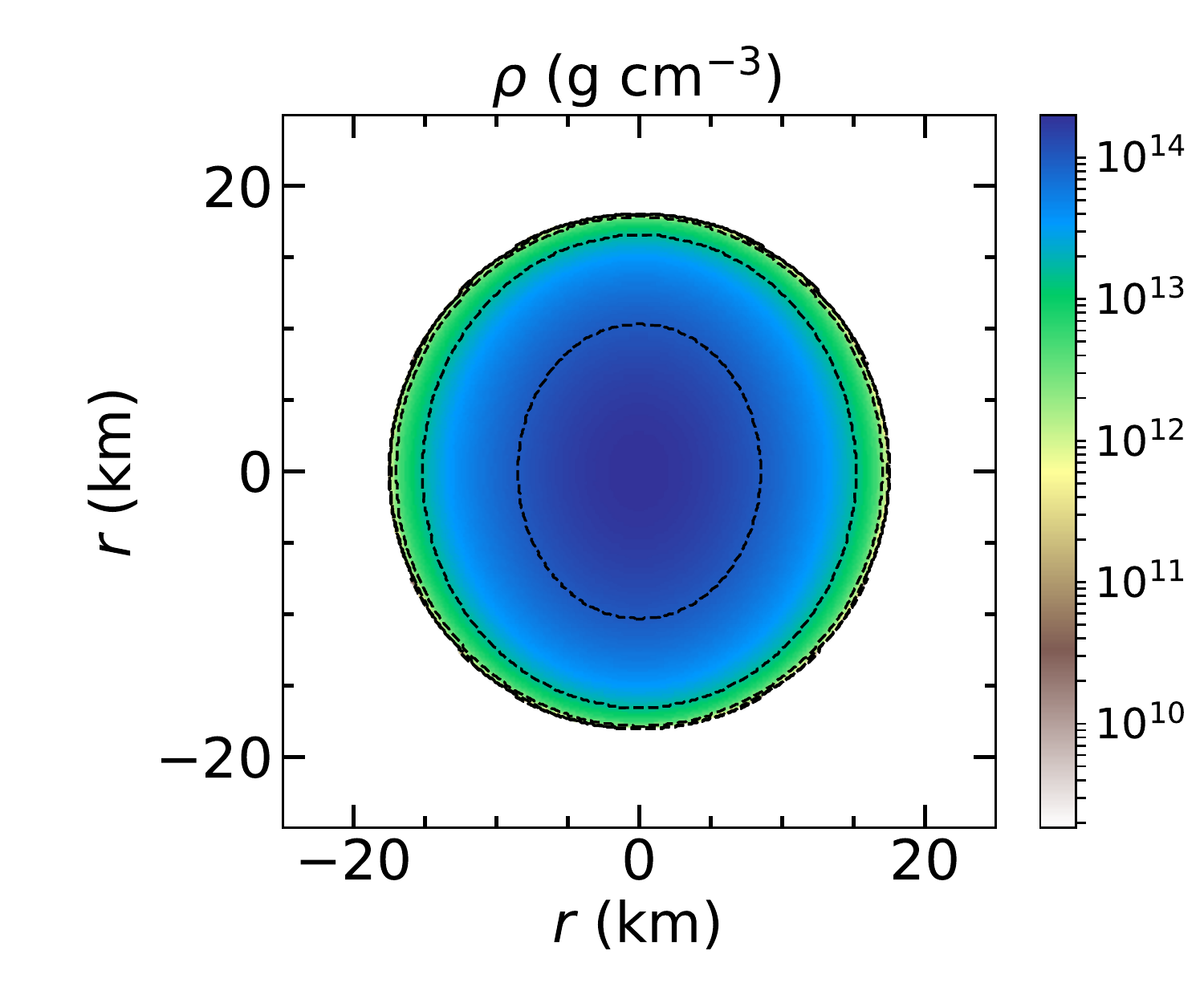}}
\subfigure[Poloidal magnetic field with $\Omega \sim 1884.9$ rad s$^{-1}$, $B_\text{max} \sim 2.1\times 10^{16}$ G, ME/GE $\sim 7.8\times10^{-4}$, KE/GE $\sim 9.1\times10^{-2}$.]{\includegraphics[scale=0.5]{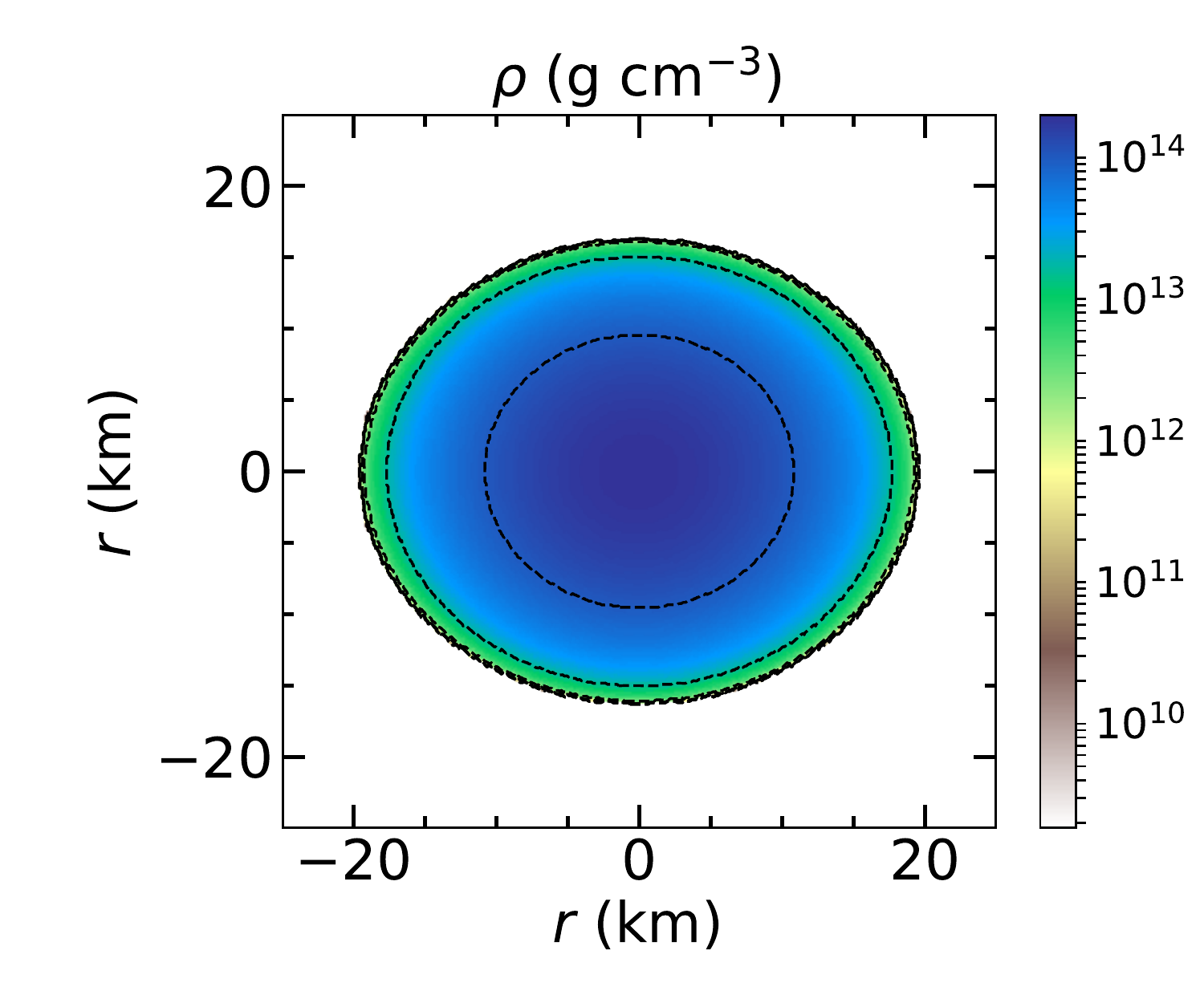}}
\caption{Density isocontours of uniformly rotating magnetized neutron star.}
\label{Fig: Neutron star}
\end{figure}

\begin{figure*}
\centering
\includegraphics[scale=0.5]{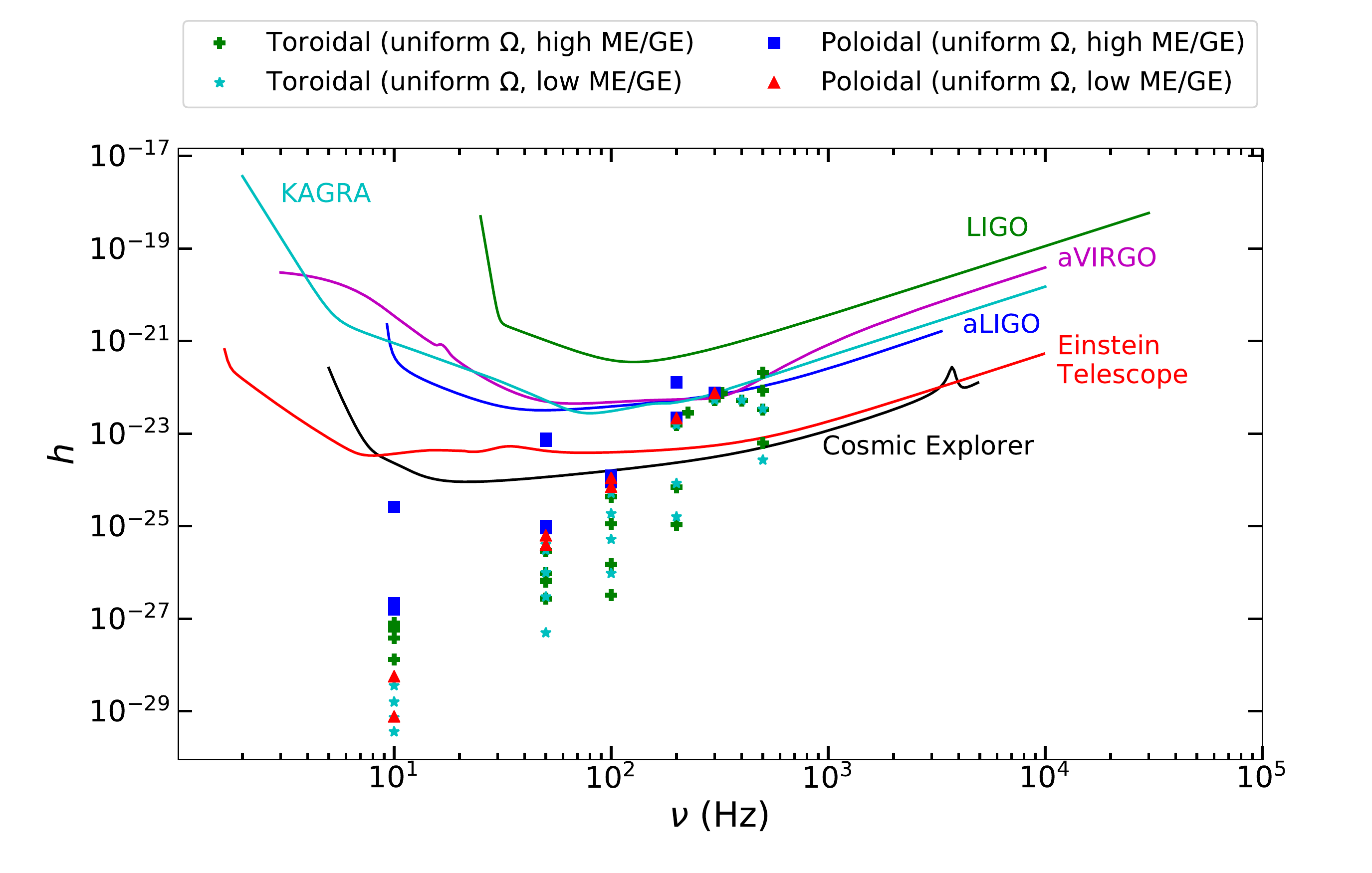}
\caption{Same as Figure \ref{Detector}, except for neutron stars, as given in Tables \ref{NS Poloidal Magnetic Field Table} and \ref{NS Toroidal Magnetic Field Table}. Here also $h = 0.0110297h_0$ with $\chi=3\degree$.}
\label{Detector_NS}
\end{figure*}

We explore the generation of continuous gravitational wave from uniformly rotating B-NSs too. Being smaller in size, neutron stars can rotate much faster than the white dwarfs and hence we choose the frequency in the range $1-500$ Hz depending on their central density. We further use the equation of state with polytropic index $2$ following \cite{2014MNRAS.439.3541P}, when {\it XNS} necessarily requires 
the equation of state in a polytropic form. 
However, if one fits the data of actual equation of states with the polytropic 
law, most of them are well fitted with the polytropic index $\sim 1.8-2.2$. 
Hence, our choice is justified. We choose purely toroidal and purely poloidal magnetic field cases separately as we have considered for the white dwarfs, 
as shown in Figure \ref{Fig: Neutron star} for two typical
examples. Various possible sets of mass and
radius of B-NSs, simulated by us, are given in Tables \ref{NS Toroidal Magnetic Field Table} and \ref{NS Poloidal Magnetic Field Table}, which are in accordance with the existing literature (see, e.g., \citealt{2012ARNPS..62..485L,2016ARA&A..54..401O,2016PhRvD..93b3501C}).
Moreover, we consider two different cases of ME/GE for each of the magnetic field geometry
(see Tables \ref{NS Toroidal Magnetic Field Table} and \ref{NS Poloidal Magnetic Field Table}) assuring stability \citep{2009MNRAS.397..763B,2013MNRAS.433.2445A,2017MNRAS.466.1330H}, which shows that neutron stars with high magnetic field emit stronger gravitational radiation than those having low magnetic field. The values of $h_0$ vs. $\nu$ for B-NSs have been shown in Figure \ref{Detector_NS}. For all the cases, we assume the distance of the neutron star from the detector to be 10 kpc. The values of $h_0$, in case of neutron stars, are tabulated in Tables \ref{NS Toroidal Magnetic Field Table}
and \ref{NS Poloidal Magnetic Field Table}. For neutron stars with toroidal magnetic field, we vary $\rho_c$ from $10^{14}$ g cm$^{-3}$ to $2\times10^{15}$ g cm$^{-3}$. However, {\it{XNS}} code could not handle poloidal magnetic field with rotation for high $\rho_c$ until recently (A. Pili, private communication) and hence we choose only two values of $\rho_c = 10^{14}$ g cm$^{-3}$ and $2\times10^{14}$ g cm$^{-3}$ in case of purely poloidal uniformly rotating neutron stars.

\begin{table*}
\centering
\caption{Uniformly rotating neutron star with toroidal magnetic field when $d = 10$ kpc with $\chi = 3\degree$.
Here $B_\text{max}$ is the maximum magnetic field close to the center of neutron star, when surface field could be much smaller.}
\label{NS Toroidal Magnetic Field Table}
\begin{tabular}{|l|l|l|l|l|l|l|l|l|l|}
\hline
$\rho_c$ (g cm$^{-3}$) & $M$ ($M_\odot$)& $R_E$ (km) & $R_P/R_E$ & $B_{\max}$ (G) & $\nu$ (Hz) & ME/GE & KE/GE & $|I_{x'x'}-I_{y'y'}|/I_{z'z'}$ & $h_0$\\
\hline\hline
	$10^{14}$ & 0.406 & 17.8 & 1.0000 & $6.832\times 10^{15}$ & 10.0 & $7.71\times10^{-4}$ & $1.60\times 10^{-4}$ & $5.71\times10^{-6}$ & $5.2013\times 10^{-26}$\\
	& 0.408 & 18.0 & 0.9901 & $6.850\times 10^{15}$ & 50.0 & $7.76\times10^{-4}$ & $4.04\times 10^{-3}$ & $1.14\times10^{-5}$ & $2.6239\times 10^{-24}$\\
	& 0.414 & 18.2 & 0.9707 & $6.909\times 10^{15}$ & 100.0 & $7.89\times10^{-4}$ & $1.65\times 10^{-2}$ & $7.04\times10^{-5}$ & $6.7539\times 10^{-23}$\\
	& 0.444 & 19.6 & 0.8552 & $6.901\times 10^{15}$ & 200.0 & $7.86\times10^{-4}$ & $7.33\times 10^{-2}$ & $3.01\times10^{-4}$ & $1.3838\times 10^{-21}$\\
	& 0.459 & 20.3 & 0.8079 & $6.920\times 10^{15}$ & 226.1 & $7.90\times10^{-4}$ & $9.85\times 10^{-2}$ & $4.01\times10^{-4}$ & $2.5643\times 10^{-21}$\\
\hline
	$2\times10^{14}$ & 0.711 & 16.6 & 1.0000 & $1.339\times 10^{16}$ & 10.0 & $7.90\times10^{-4}$ & $7.94\times 10^{-5}$ & $5.91\times10^{-6}$ & $7.2360\times 10^{-26}$\\
	& 0.712 & 16.6 & 1.0000 & $1.341\times 10^{16}$ & 50.0 & $7.92\times10^{-4}$ & $1.99\times 10^{-3}$ & $2.80\times10^{-6}$ & $8.6017\times 10^{-25}$\\
	& 0.718 & 16.7 & 0.9788 & $1.336\times 10^{16}$ & 100.0 & $7.86\times10^{-4}$ & $8.05\times 10^{-3}$ & $3.14\times10^{-5}$ & $3.9250\times 10^{-23}$\\
	& 0.784 & 18.3 & 0.8454 & $1.337\times 10^{16}$ & 300.0 & $7.89\times10^{-4}$ & $8.29\times 10^{-2}$ & $3.41\times10^{-4}$ & $4.7777\times 10^{-21}$\\
	& 0.799 & 18.9 & 0.8028 & $1.318\times 10^{16}$ & 323.0 & $7.68\times10^{-4}$ & $9.90\times 10^{-2}$ & $3.96\times10^{-5}$ & $6.7577\times 10^{-21}$\\
\hline
	$5\times10^{14}$ & 1.261 & 13.9 & 1.0000 & $3.087\times 10^{16}$ & 10.0 & $7.93\times10^{-4}$ & $3.10\times 10^{-5}$ & $5.84\times10^{-6}$ & $6.6069\times 10^{-26}$\\
	& 1.262 & 13.9 & 1.0000 & $3.068\times 10^{16}$ & 50.0 & $7.83\times10^{-4}$ & $7.77\times 10^{-4}$ & $2.17\times10^{-6}$ & $6.1344\times 10^{-25}$\\
	& 1.265 & 13.9 & 1.0000 & $3.071\times 10^{16}$ & 100.0 & $7.85\times10^{-4}$ & $3.12\times 10^{-3}$ & $8.98\times10^{-6}$ & $1.0232\times 10^{-23}$\\
	& 1.332 & 14.8 & 0.8922 & $3.056\times 10^{16}$ & 400.0 & $7.84\times10^{-4}$ & $5.43\times 10^{-2}$ & $2.30\times10^{-4}$ & $4.6925\times 10^{-21}$\\
	& 1.393 & 15.5 & 0.8171 & $3.045\times 10^{16}$ & 516.9 & $7.84\times10^{-4}$ & $9.76\times 10^{-2}$ & $3.88\times10^{-4}$ & $1.4643\times 10^{-20}$\\
\hline
	$10^{15}$ & 1.613 & 11.3 & 1.0000 & $5.493\times 10^{16}$ & 10.0 & $7.86\times10^{-4}$ & $1.51\times 10^{-5}$ & $5.12\times10^{-6}$ & $3.4354\times 10^{-26}$\\
	& 1.613 & 11.3 & 1.0000 & $5.493\times 10^{16}$ & 50.0 & $7.86\times10^{-4}$ & $3.78\times 10^{-4}$ & $3.39\times10^{-6}$ & $5.6965\times 10^{-25}$\\
	& 1.615 & 11.3 & 1.0000 & $5.494\times 10^{16}$ & 100.0 & $7.86\times10^{-4}$ & $1.51\times 10^{-3}$ & $1.99\times10^{-5}$ & $1.3414\times 10^{-24}$\\
	& 1.621 & 11.3 & 0.9843 & $5.499\times 10^{16}$ & 200.0 & $7.89\times10^{-4}$ & $6.10\times 10^{-3}$ & $2.33\times10^{-5}$ & $6.3106\times 10^{-23}$\\
	& 1.668 & 11.6 & 0.9237 & $5.473\times 10^{16}$ & 500.0 & $7.89\times10^{-4}$ & $4.01\times 10^{-2}$ & $1.69\times10^{-4}$ & $3.0187\times 10^{-21}$\\
	& 1.690 & 16.7 & 0.9048 & $6.353\times 10^{17}$ & 500.0 & 0.1399 & $4.69\times 10^{-2}$ & $7.75\times10^{-4}$ & $1.8769\times 10^{-19}$\\
\hline
	$2\times10^{15}$ & 1.712 & 8.6 & 1.0000 & $9.325\times 10^{16}$ & 10.0 & $7.88\times10^{-4}$ & $7.36\times 10^{-6}$ & $4.21\times10^{-6}$ & $1.1923\times 10^{-26}$\\
	& 1.713 & 8.6 & 1.0000 & $9.324\times 10^{16}$ & 50.0 & $7.88\times10^{-4}$ & $1.84\times 10^{-4}$ & $3.44\times10^{-6}$ & $2.4359\times 10^{-25}$\\
	& 1.713 & 8.6 & 1.0000 & $9.324\times 10^{16}$ & 100.0 & $7.89\times10^{-4}$ & $7.37\times 10^{-4}$ & $1.03\times10^{-6}$ & $2.9285\times 10^{-25}$\\
	& 1.716 & 8.8 & 1.0000 & $9.323\times 10^{16}$ & 200.0 & $7.89\times10^{-4}$ & $2.96\times 10^{-3}$ & $8.60\times10^{-6}$ & $9.7596\times 10^{-24}$\\
	& 1.735 & 8.8 & 0.9596 & $9.269\times 10^{16}$ & 500.0 & $7.86\times10^{-4}$ & $1.89\times 10^{-2}$ & $7.92\times10^{-5}$ & $5.7122\times 10^{-22}$\\
	& 1.786 & 11.8 & 0.9849 & $1.119\times 10^{18}$ & 500.0 & 0.1400 & $2.15\times 10^{-2}$ & $7.72\times10^{-4}$ & $7.6570\times 10^{-20}$\\
\hline\hline
	$10^{14}$ & 0.406 & 17.8 & 1.0000 & $1.038\times 10^{15}$ & 10.0 & $1.78\times10^{-5}$ & $1.60\times 10^{-4}$ & $3.55\times10^{-7}$ & $3.2372\times 10^{-27}$\\
	& 0.408 & 18.0 & 0.9901 & $1.041\times 10^{15}$ & 50.0 & $1.78\times10^{-5}$ & $4.01\times 10^{-3}$ & $1.58\times10^{-5}$ & $3.6394\times 10^{-24}$\\
	& 0.414 & 18.2 & 0.9707 & $1.050\times 10^{15}$ & 100.0 & $1.79\times10^{-5}$ & $1.62\times 10^{-2}$ & $7.64\times10^{-5}$ & $7.3383\times 10^{-23}$\\
	& 0.445 & 19.6 & 0.8552 & $1.036\times 10^{15}$ & 200.0 & $1.64\times10^{-5}$ & $6.82\times 10^{-2}$ & $3.06\times10^{-4}$ & $1.4061\times 10^{-21}$\\
\hline
	$2\times10^{14}$ & 0.711 & 16.6 & 1.0000 & $1.055\times 10^{15}$ & 10.0 & $4.89\times10^{-6}$ & $7.94\times 10^{-5}$ & $1.17\times10^{-7}$ & $1.4329\times 10^{-27}$\\
	& 0.713 & 16.6 & 1.0000 & $1.057\times 10^{15}$ & 50.0 & $4.89\times10^{-6}$ & $1.99\times 10^{-3}$ & $8.91\times10^{-6}$ & $2.7422\times 10^{-24}$\\
	& 0.718 & 16.7 & 0.9788 & $1.060\times 10^{15}$ & 100.0 & $4.90\times10^{-6}$ & $7.98\times 10^{-3}$ & $3.74\times10^{-5}$ & $4.6802\times 10^{-23}$\\
	& 0.784 & 18.3 & 0.8357 & $1.008\times 10^{15}$ & 300.0 & $4.13\times10^{-6}$ & $7.64\times 10^{-2}$ & $3.45\times10^{-4}$ & $4.8414\times 10^{-21}$\\
\hline
	$5\times10^{14}$ & 1.262 & 13.9 & 1.0000 & $1.000\times 10^{15}$ & 10.0 & $8.28\times10^{-7}$ & $3.10\times 10^{-5}$ & $2.88\times10^{-8}$ & $3.2551\times 10^{-28}$\\
	& 1.263 & 13.9 & 1.0000 & $1.000\times 10^{15}$ & 50.0 & $8.28\times10^{-7}$ & $7.75\times 10^{-4}$ & $3.15\times10^{-6}$ & $8.9211\times 10^{-25}$\\
	& 1.266 & 13.9 & 0.9873 & $1.000\times 10^{15}$ & 100.0 & $8.28\times10^{-7}$ & $3.10\times 10^{-3}$ & $1.48\times10^{-5}$ & $1.6869\times 10^{-23}$\\
	& 1.333 & 14.8 & 0.8802 & $1.021\times 10^{15}$ & 400.0 & $8.28\times10^{-7}$ & $5.14\times 10^{-2}$ & $2.35\times10^{-4}$ & $4.8061\times 10^{-21}$\\
\hline
	$10^{15}$ & 1.613 & 11.3 & 1.0000 & $1.000\times 10^{15}$ & 10.0 & $2.59\times10^{-7}$ & $1.51\times 10^{-5}$ & $9.86\times10^{-8}$ & $6.6142\times 10^{-28}$\\
	& 1.614 & 11.3 & 1.0000 & $1.000\times 10^{15}$ & 50.0 & $2.59\times10^{-7}$ & $3.77\times 10^{-4}$ & $1.61\times10^{-6}$ & $2.6955\times 10^{-25}$\\
	& 1.615 & 11.3 & 1.0000 & $1.000\times 10^{15}$ & 100.0 & $2.59\times10^{-7}$ & $1.51\times 10^{-3}$ & $7.01\times10^{-6}$ & $4.7125\times 10^{-24}$\\
	& 1.621 & 11.3 & 0.9843 & $1.000\times 10^{15}$ & 200.0 & $2.58\times10^{-7}$ & $6.05\times 10^{-3}$ & $2.81\times10^{-5}$ & $7.6173\times 10^{-23}$\\
	& 1.669 & 11.6 & 0.9237 & $1.006\times 10^{15}$ & 500.0 & $2.55\times10^{-7}$ & $3.85\times 10^{-2}$ & $1.73\times10^{-4}$ & $3.0893\times 10^{-21}$\\
\hline
	$2\times10^{15}$ & 1.713 & 8.6 & 1.0000 & $1.040\times 10^{15}$ & 50.0 & $9.79\times10^{-8}$ & $1.84\times 10^{-4}$ & $6.36\times10^{-7}$ & $4.4945\times 10^{-26}$\\
	& 1.713 & 8.6 & 1.0000 & $1.040\times 10^{15}$ & 100.0 & $9.78\times10^{-8}$ & $7.35\times 10^{-4}$ & $3.05\times10^{-6}$ & $8.6329\times 10^{-25}$\\
	& 1.716 & 8.6 & 1.0000 & $1.040\times 10^{15}$ & 200.0 & $9.77\times10^{-8}$ & $2.94\times 10^{-3}$ & $1.27\times10^{-5}$ & $1.4416\times 10^{-23}$\\
	& 1.735 & 8.8 & 0.9596 & $1.039\times 10^{15}$ & 500.0 & $9.67\times10^{-8}$ & $1.85\times 10^{-2}$ & $8.29\times10^{-5}$ & $2.4411\times 10^{-22}$\\
\hline
\end{tabular}
\end{table*}

\begin{table*}
\centering
\caption{Uniformly rotating neutron star with poloidal magnetic field when $d = 10$ kpc with $\chi = 3\degree$.
Here $B_\text{max}$ is the maximum magnetic field at the center of neutron star, when surface field could be much smaller.}
\label{NS Poloidal Magnetic Field Table}
\begin{tabular}{|l|l|l|l|l|l|l|l|l|l|}
\hline
$\rho_c$ (g cm$^{-3}$) & $M$ ($M_\odot$)& $R_E$ (km) & $R_P/R_E$ & $B_{\max}$ (G) & $\nu$ (Hz) & ME/GE & KE/GE & $|I_{x'x'}-I_{y'y'}|/I_{z'z'}$ & $h_0$\\
\hline\hline
	$10^{14}$ & 0.491 & 18.9 & 0.9906 & $1.088\times 10^{16}$ & 10.0 & $7.70\times10^{-4}$ & $1.75\times 10^{-3}$ & $1.18\times10^{-5}$ & $1.4385\times 10^{-25}$\\
	& 0.493 & 18.9 & 0.9906 & $1.092\times 10^{16}$ & 50.0 & $7.89\times10^{-4}$ & $1.14\times 10^{-2}$ & $2.97\times10^{-5}$ & $9.1324\times 10^{-24}$\\
	& 0.501 & 19.2 & 0.9631 & $1.076\times 10^{16}$ & 100.0 & $7.83\times10^{-4}$ & $3.01\times 10^{-2}$ & $8.86\times10^{-5}$ & $1.1326\times 10^{-22}$\\
	& 0.539 & 20.6 & 0.8541 & $1.027\times 10^{16}$ & 200.0 & $7.54\times10^{-4}$ & $9.65\times 10^{-2}$ & $3.26\times10^{-4}$ & $1.9999\times 10^{-21}$\\
	& 0.738 & 21.7 & 0.6245 & $1.440\times 10^{17}$ & 10.0 & 0.1097 & $1.45\times 10^{-3}$ & $8.92\times10^{-4}$ & $2.3552\times 10^{-23}$\\
	& 0.757 & 21.9 & 0.6113 & $1.471\times 10^{17}$ & 50.0 & 0.1136 & $1.00\times 10^{-2}$ & $9.11\times10^{-4}$ & $6.2795\times 10^{-22}$\\
\hline
	$2\times10^{14}$ & 0.898 & 17.6 & 0.9899 & $2.151\times 10^{16}$ & 10.0 & $7.85\times10^{-4}$ & $7.92\times 10^{-4}$ & $1.16\times10^{-5}$ & $1.9477\times 10^{-25}$\\
	& 0.900 & 17.6 & 0.9899 & $2.093\times 10^{16}$ & 50.0 & $7.49\times10^{-4}$ & $5.22\times 10^{-3}$ & $1.93\times10^{-5}$ & $8.1151\times 10^{-24}$\\
	& 0.907 & 17.8 & 0.9801 & $2.106\times 10^{16}$ & 100.0 & $7.67\times10^{-4}$ & $1.38\times 10^{-2}$ & $4.79\times10^{-5}$ & $8.2153\times 10^{-23}$\\
	& 0.993 & 19.6 & 0.8281 & $2.056\times 10^{16}$ & 300.0 & $7.76\times10^{-4}$ & $9.08\times 10^{-2}$ & $3.63\times10^{-4}$ & $6.9602\times 10^{-21}$\\
	& 1.325 & 19.9 & 0.6178 & $2.658\times 10^{17}$ & 200.0 & 0.1000 & $3.80\times 10^{-2}$ & $8.94\times10^{-4}$ & $1.1603\times 10^{-20}$\\
	& 1.329 & 19.6 & 0.6199 & $2.896\times 10^{17}$ & 50.0 & 0.1163 & $3.96\times 10^{-3}$ & $8.99\times10^{-4}$ & $7.1393\times 10^{-22}$\\
\hline\hline
	$10^{14}$ & 0.490 & 18.9 & 1.0000 & $1.004\times 10^{15}$ & 10.0 & $6.54\times10^{-6}$ & $1.75\times 10^{-3}$ & $4.22\times10^{-7}$ & $5.1038\times 10^{-27}$\\
	& 0.492 & 18.9 & 0.9906 & $1.008\times 10^{15}$ & 50.0 & $6.64\times10^{-6}$ & $1.13\times 10^{-2}$ & $1.86\times10^{-5}$ & $5.6962\times 10^{-24}$\\
	& 0.500 & 19.2 & 0.9631 & $1.007\times 10^{15}$ & 100.0 & $6.66\times10^{-6}$ & $2.92\times 10^{-2}$ & $7.82\times10^{-5}$ & $9.9460\times 10^{-23}$\\
	& 0.537 & 20.6 & 0.8541 & $1.059\times 10^{15}$ & 200.0 & $7.32\times10^{-6}$ & $8.79\times 10^{-2}$ & $3.18\times10^{-4}$ & $1.9420\times 10^{-21}$\\
\hline
	$2\times10^{14}$ & 0.896 & 17.6 & 1.0000 & $1.042\times 10^{15}$ & 10.0 & $1.84\times10^{-6}$ & $7.91\times 10^{-4}$ & $4.15\times10^{-8}$ & $6.9342\times 10^{-28}$\\
	& 0.898 & 17.6 & 1.0000 & $1.044\times 10^{15}$ & 50.0 & $1.85\times10^{-6}$ & $5.20\times 10^{-3}$ & $8.52\times10^{-6}$ & $3.5748\times 10^{-24}$\\
	& 0.905 & 17.8 & 0.9801 & $1.050\times 10^{15}$ & 100.0 & $1.88\times10^{-6}$ & $1.36\times 10^{-2}$ & $3.73\times10^{-5}$ & $6.3668\times 10^{-23}$\\
	& 0.991 & 19.6 & 0.8281 & $1.042\times 10^{15}$ & 300.0 & $1.83\times10^{-6}$ & $8.32\times 10^{-2}$ & $3.55\times10^{-4}$ & $6.7803\times 10^{-21}$\\
\hline
\end{tabular}
\end{table*}

Interestingly, there is no detection of continuous gravitational wave from neutron stars in LIGO so far and it is well in accordance with Figure \ref{Detector_NS}. If any of them is detected in future by aLIGO, aVIRGO, Einstein Telescope, Cosmic Explorer etc., depending on its distance from the earth, then we can make a prediction of the magnetic field in neutron stars. Nevertheless, a fast-spinning neutron star with a strong field would not sustain its fast rotation for long due to its efficient spin-down luminosity. Hence, in practice they are difficult to detect, unless captured at the very birth stage \citep{2018MNRAS.480.1353D}. 
Moreover, similar analysis of GW for other exotic stars, e.g. quark stars \citep{2006JPhG...32.1081M} etc., is
expected to offer to constrain their various properties including the mass-radius relation, which is useful to carry out in future.

%---------------------------------------------------------------------------------------------------------------------
\section{Luminosities due to gravitational radiation and electromagnetic radiation}\label{gwem}
Since the B-WDs considered here have magnetic field and rotation both, they may behave as a rotating dipole. Therefore, they must possess luminosity due to dipole radiation along with gravitational radiation which is quadrupolar in nature. In other words, B-WDs have electromagnetic counterparts. The luminosity due to gravitational radiation is given by \citep{2009igr..book.....R}
\begin{align}
L_\text{GW} &= \frac{G}{c^5}\left\langle\dddot{Q}_{ij} \dddot{Q}_{ij}\right\rangle \\
&= \frac{G\Omega^6}{5c^5} \epsilon^2 I_{xx}^2 \sin^2\chi (2\cos^2\chi - \sin^2\chi)^2 \nonumber\\ &\left\{\frac{1}{4}\cos^2\chi \sin^2 i (1+\cos^2i) + \sin^2\chi (1+6 \cos^2i+ \cos^4i) \right\}.
\end{align}
The detailed derivation of this formula is given in Appendix \ref{appendix2}. It is evident from this formula that $L_\text{GW}$ is directly proportional to $\sin^2\chi$, which also verifies that there will be no gravitational radiation if the magnetic and rotation axes are aligned. On the other hand, in the Newtonian limit, the luminosity due to electromagnetic dipole radiation is given by \citep{2016JCAP...05..007M}
\begin{align}
L_\text{EM} = \frac{4\Omega^4\sin^2\chi}{5c^3}|m|^2,
\end{align}
where $m$ is the magnetic dipole moment, which is related to the surface magnetic field at the pole as
\begin{align}
B_s = \frac{2|m|}{R^3}.
\end{align}
The exact formula for luminosity due to electromagnetic dipole radiation considering general relativistic (GR) effect was obtained by \cite{2004MNRAS.352.1161R,2016MNRAS.459.4144R}. However, in case of white dwarfs, GR effect in spin-down luminosity is not very significant (as it is altered with a small factor and the order of magnitude of the luminosity remains the same), we just follow the Newtonian formula for our calculations.
Now, if the body has a rotational period $P$ which is expected to be changing with time as
$\dot{P}$, then
\begin{align}
B_s = \sqrt{\frac{5c^3 I_{z'z'} P \dot{P}}{4 \pi^2 R^6 \sin^2\chi}}\, \text{G}.
\end{align}
Therefore, the luminosity due to dipole radiation reduces to
\begin{align}
L_\text{EM} = 4\pi^2I_{z'z'} \frac{\dot{P}}{P^3}.
\end{align}
Table \ref{Luminosity_table} shows $L_\text{GW}$ and $L_\text{EM}$ for a few typical cases for white dwarfs, assuming $\dot{P}=10^{-15}$ Hz s$^{-1}$. It is found that the luminosity ranges for electromagnetic and gravitational radiations are different from each other. This will be another unique way of separating B-WDs from regular white dwarfs. While regular non-magnetized or weakly magnetized white dwarfs do not have any electromagnetic counterparts, $L_\text{EM}$ for B-WDs could be above $10^{34}$ ergs s$^{-1}$, as given in Table \ref{Luminosity_table}, which are already observed in many magnetized compact sources including white dwarfs \citep{2016Natur.537..374M, 2013ApJ...770...65R, 2014ApJ...784...37D, 2014ApJ...786...62S, 2016JCAP...05..007M}. Nevertheless, with increasing $\dot{P}$, $L_\text{EM}$ as well as $B_s$ increase. Hence, in some cases, $B_s$ may turn out to be well above $10^9$ G, above the maximum $B_s$ of white dwarfs currently inferred from observation. Therefore, GW astronomy may be quite useful to identify or to rule out such predicted B-WDs. Moreover, the thermal time scale (also known as the Kelvin-Helmholtz time scale) is defined as
\begin{equation}
\tau_{KH} = \frac{GM^2}{RL},
\end{equation}
where $M$, $R$ and $L$ are respectively the mass, radius and luminosity of the body. Substituting the values of $L_\text{GW}$ from Table \ref{Luminosity_table}, we obtain that $\tau_{KH} \sim 10^{7-8}$ years.

\begin{table*}
\centering
\caption{$L_\text{GW}$ and $L_\text{EM}$ for white dwarfs considering $\dot{P}=10^{-15}$ Hz s$^{-1}$
and $\chi=3\degree$. $B_s$ is the surface magnetic field at the pole.}
\label{Luminosity_table}
\begin{tabular}{|l|l|l|l|l|l|l|}
\hline
	$M$ ($M_\odot$) & $R$ (km) & $I_{z'z'}$ (g cm$^2$) & $P$ (s) & $B_s$ (G)& $L_\text{GW}$ (ergs s$^{-1}$) & $L_\text{EM}$ (ergs s$^{-1}$)\\
\hline\hline
	1.420 & 1718.8 & $5.17\times10^{48}$ & 1.5 & $6.12\times10^{8}$ & $2.91\times10^{35}$ & $5.50\times10^{34}$\\
	1.640 & 1120.7 & $6.13\times10^{48}$ & 2.0 & $2.78\times10^{9}$ & $3.46\times10^{36}$ & $3.03\times10^{34}$\\
	1.702 & 1027.2 & $1.92\times10^{48}$ & 3.1 & $4.52\times10^{9}$ & $3.17\times10^{35}$ & $8.23\times10^{33}$\\
\hline
\end{tabular}
\end{table*}

%=====================================================================================================================
\section{Conclusions}\label{conclusion}

After the discovery of gravitational wave from the merger events, the search for continuous gravitational wave has been a great interest in the scientific community. Undoubtedly, compact sources like neutron stars and white dwarfs are good candidates for this purpose. Due to smaller size of the neutron stars, they can rotate much faster than the white dwarfs, resulting in generation of stronger gravitation radiation and may be detected by aLIGO, aVIRGO, Einstein Telescope etc. On the other hand, although white dwarfs are bigger in size and cannot rotate as fast as neutron stars, yet they can also emit significant amount of gravitational radiation, provided they possess non-zero quadrupole moment. White dwarfs are usually closer to Earth and $h_0 \propto 1/d$, hence the strength will be higher. Moreover, because of the bigger size of the white dwarf, its moment of inertia is higher compared to that of neutron star as both of them possess similar mass; and since $h_0\propto \epsilon I_{xx}$, the strength could also be higher. We argue that, in future, these highly magnetized rotating white dwarfs, namely B-WDs, can prominently be detected by LISA, eLISA, ALIA, DECIGO and BBO detectors.

The possible existence super-Chandrasekhar white dwarfs as inferred 
from observations has stimulated astronomers a lot in the past decade.
However, it has, so far, only been detected indirectly from the lightcurve of over-luminous peculiar type Ia supernovae \citep{2006Natur.443..308H, 2010ApJ...713.1073S}. As we have discussed in section \ref{Introduction}, many theories have been proposed to explain the violation of Chandrasekhar mass-limit. The detection of continuous gravitational wave from white dwarfs or B-WDs will confirm these objects directly. We have used the {\it{XNS}} code to determine the structure of white dwarfs as well as neutron stars. Although {\it{XNS}} code has a couple of limitations such as the requirement to supply a polytropic equation of state and the implicit assumption of $\chi=0$, we overcome these shortcomings with the following assumptions. First, we supply the polytropic equation of state in such a way that it almost represents the actual mass-radius relation of the compact objects. Second, if the magnetic field and rotation axes are aligned to each other, the object does not radiate any gravitational radiation and, hence, we throughout assume small angle approximation to avoid the ambiguity in the structure of the object. However, had we run an efficient code with appropriately chosen $\chi$, we would have been able to generate gravitational wave with much higher strength as the strength monotonically increases with the angle $\chi$ and it becomes maximum at $\chi=90\degree$.

\section*{acknowledgments}
The authors would like to thank A. Gopakumar of TIFR, Mumbai, for useful discussion and suggestion during compilation of the work. We also thank Sanjit Mitra of IUCAA, Pune, for providing some updated information in gravitational wave astronomy. S. K. thanks Soheb Mandhai of University of Leicester and Adam Pound of University of Southampton for discussion about the sensitivity curves and confusion noise. We also thank Sathyawageeswar Subramanian of University of Cambridge for helping with use of {\it{XNS}} code for white dwarfs and Upasana Das of NORDITA, Stockholm, for providing useful references. S. K. would like to thank Timothy Brandt of University of California, Santa Barbara, for the useful discussion about the Kelvin-Helmholtz time-scale. B. M. would like to thank Tom Marsh of University of Warwick and Tomasz Bulik of Nicolaus Copernicus Astronomical Center (CAMK) for discussion in the conference ``Compact White Dwarf Binaries'', Yerevan, Armenia. Finally, thanks are due to the anonymous referee for thorough reading the manuscript and comments which have helped to improve the presentation of the work. The work was partially supported by a project supported by Department of Science and Technology (DST), India, with Grant No. DSTO/PPH/BMP/1946 (EMR/2017/001226).

%%%%%%%%%%%%%%%%%%%%%%%%%%%%%%%%%%%%%%%%%%%%%%%%%%

%%%%%%%%%%%%%%%%%%%% REFERENCES %%%%%%%%%%%%%%%%%%

\bibliographystyle{mnras}
\bibliography{mypaper2}

%%%%%%%%%%%%%%%%%%%%%%%%%%%%%%%%%%%%%%%%%%%%%%%%%%

%%%%%%%%%%%%%%%%% APPENDICES %%%%%%%%%%%%%%%%%%%%%

\appendix
\section{Derivation of the amplitude of GW}\label{appendix1}

The gravitational wave amplitude $h_0$ is given by equation \eqref{grav_wave_amplitude}, which is
\begin{equation*}
h_0 = -\frac{6G}{c^4}Q_{z'z'}\frac{\Omega^2}{d}.
\end{equation*}
Substituting equations \eqref{new moment of inertia} and \eqref{quadrupole moment} in the above equation, we obtain
\begin{align*}
h_0 &= -\frac{6G}{c^4}\frac{\Omega^2}{d}\Big(-I_{z'z'}+\frac{1}{3}(I_{x'x'}+I_{y'y'}+I_{z'z'})\Big)\\
&= -\frac{2G}{c^4}\frac{\Omega^2}{d}(I_{x'x'}+I_{y'y'}-2I_{z'z'})\\
&= -\frac{2G}{c^4}\frac{\Omega^2}{d}\Big(I_{xx}\cos^2\chi+I_{zz}\sin^2\chi+I_{yy}-2(I_{xx}\sin^2\chi+I_{zz}\cos^2\chi)\Big)\\
&= -\frac{2G}{c^4}\frac{\Omega^2}{d}\Big(I_{xx}(\cos^2\chi+1-2\sin^2\chi)+I_{zz}(\sin^2\chi-2\cos^2\chi)\Big)\\
&= -\frac{2G}{c^4}\frac{\Omega^2}{d}\Big(I_{xx}(2\cos^2\chi-\sin^2\chi)-I_{zz}(2\cos^2\chi-\sin^2\chi)\Big)\\
&= -\frac{2G}{c^4}\frac{\Omega^2}{d}(I_{xx}-I_{zz})(2\cos^2\chi-\sin^2\chi)\\
&= \frac{2G}{c^4}\frac{\Omega^2\epsilon I_{xx}}{d}(2\cos^2\chi-\sin^2\chi).
\end{align*}
Here we use $I_{xx}=I_{yy}$, as the object is symmetric about $z-$axis and define $\epsilon = (I_{zz}-I_{xx})/I_{xx}$.

\section{Derivation of the formula for luminosity due to gravitational wave}\label{appendix2}

The relation between the quadrupolar moment and gravitational wave strength is given by
\begin{align}\label{Eq: A1}
h_{ij} = \frac{2G}{c^4 d} \ddot{Q}_{ij},
\end{align}
where $d$ is the distance of the source from the detector. Moreover, the relation between GW luminosity and quadrupolar moment is \citep{2009igr..book.....R}
\begin{align}\label{Eq: A2}
L_\text{GW} = \frac{G}{5c^5}\left\langle\dddot{Q}_{ij} \dddot{Q}_{ij}\right\rangle.
\end{align}
Combining these two equations \eqref{Eq: A1} and \eqref{Eq: A2}, we obtain
\begin{align}\label{Eq: LGW}
L_\text{GW} = \frac{c^3 d^2}{20G}\left\langle\dot{h}_{ij} \dot{h}_{ij}\right\rangle.
\end{align}
Moreover, using the relation $\left\langle\dot{h}_{ij} \dot{h}_{ij}\right\rangle = 2[\langle\dot{h}_{+}^2\rangle + \langle\dot{h}_{\times}^2\rangle]$ with $h_+$ and $h_\times$ being the two polarizations of GW, equation \eqref{Eq: LGW} reduces to
\begin{align}\label{Eq: GW}
L_\text{GW} &= \frac{c^3 d^2}{10G}[\langle\dot{h}_{+}^2\rangle + \langle\dot{h}_{\times}^2\rangle].
\end{align}
Now from the relations of equation \eqref{gravitational polarization}, the polarizations of GW are given by
\begin{equation}
\begin{aligned}
h_+ &= h_0\sin\chi\Bigg[\frac{1}{2}\cos i \sin i\cos\chi\cos\Omega t-\frac{1+\cos^2i}{2}\sin\chi\cos2\Omega t\Bigg],\\
h_\times &= h_0\sin\chi\Bigg[\frac{1}{2}\sin i\cos\chi\sin\Omega t-\cos i\sin\chi\sin2\Omega t\Bigg],
\end{aligned}
\end{equation}
with the amplitude given by
\begin{equation}
h_0 = \frac{2G}{c^4}\frac{\Omega^2\epsilon I_{xx}}{d}(2\cos^2\chi-\sin^2\chi).
\end{equation}
Therefore the time derivative of the above polarizations are given by
\begin{equation}
\begin{aligned}
\dot{h}_+ &= h_0\sin\chi\Bigg[-\frac{\Omega}{2}\cos i \sin i\cos\chi\sin\Omega t-(1+\cos^2i) \Omega\sin\chi\sin2\Omega t\Bigg],\\
\dot{h}_\times &= h_0\sin\chi\Bigg[\frac{\Omega}{2}\sin i\cos\chi\cos\Omega t-2\Omega\cos i\sin\chi\cos2\Omega t\Bigg].
\end{aligned}
\end{equation}
Hence the average values of $\dot{h}^2_+$ and $\dot{h}^2_\times$ are given by
\begin{equation}\label{Eq: A8}
\begin{aligned}
\langle\dot{h}^2_+\rangle &= h_0^2\sin^2\chi ~\Omega^2\Bigg[\frac{1}{4}\cos^2 i \sin^2 i\cos^2\chi\frac{1}{2}+(1+\cos^2i)^2 \sin^2\chi \frac{1}{2}\Bigg],\\
\langle\dot{h}^2_\times\rangle &= h^2_0\sin^2\chi ~\Omega^2\Bigg[\frac{1}{4}\sin^2 i\cos^2\chi\frac{1}{2}+4\cos^2 i \sin^2\chi \frac{1}{2}\Bigg].
\end{aligned}
\end{equation}
Substituting expressions from equations \eqref{Eq: A8} in equation \eqref{Eq: GW}, we obtain
\begin{align}
L_\text{GW} &= \frac{G\Omega^6}{5c^5} \epsilon^2 I_{xx}^2 \sin^2\chi (2\cos^2\chi - \sin^2\chi)^2 \nonumber\\ &\left\{\frac{1}{4}\cos^2\chi \sin^2 i (1+\cos^2i) + \sin^2\chi (1+6 \cos^2i+ \cos^4i) \right\}.
\end{align}
This is the exact expression for the gravitational wave luminosity of an isolated rotating white dwarf.

%If you want to present additional material which would interrupt the flow of the main paper,
%it can be placed in an Appendix which appears after the list of references.

%%%%%%%%%%%%%%%%%%%%%%%%%%%%%%%%%%%%%%%%%%%%%%%%%%

% Don't change these lines
\bsp	% typesetting comment
\label{lastpage}
\end{document}